\documentclass[prd,10pt,twocolumn,tightenlines,superscriptaddress,numerical,notitlepage,showpacs,amsmath,amssymb,amsfonts,aps,nofootinbib,longbibliography,floatfix]{revtex4-1}

\usepackage{combelow}
\usepackage{dsfont}
\usepackage{amsmath,amssymb,mathtools,bm,bbm,braket,slashed}
\usepackage{tikz}
\usepackage{cancel}
\usepackage[caption=false]{subfig}

\usepackage{xcolor}
\usepackage{ulem}
\usepackage{hyperref}
\hypersetup{colorlinks=true, citecolor=blue, urlcolor=blue, citecolor=red}

\definecolor{purple}{rgb}{0.8,0,0.6}

\definecolor{battleshipgrey}{rgb}{0.2, 0.52, 0.51}
\definecolor{darkgreen}{rgb}{0.12, 0.5, 0.17}

\begin{document}

\title{Inhibition of splitting of the chiral and deconfinement transition due to rotation in QCD: the phase diagram of linear sigma model coupled to Polyakov loop}

\author{Pracheta Singha}
\affiliation{Department of Physics, West University of Timi\cb{s}oara,  Bd.~Vasile P\^arvan 4, Timi\cb{s}oara 300223, Romania}

\author{Victor E. Ambru\cb{s}}
\affiliation{Department of Physics, West University of Timi\cb{s}oara,  Bd.~Vasile P\^arvan 4, Timi\cb{s}oara 300223, Romania}

\author{Maxim N. Chernodub}
\affiliation{Institut Denis Poisson UMR 7013, Universit\'e de Tours, 37200 France}
\affiliation{Department of Physics, West University of Timi\cb{s}oara,  Bd.~Vasile P\^arvan 4, Timi\cb{s}oara 300223, Romania}

\date{\today}

\begin{abstract}
We discuss the effect of rigid rotation on the critical temperatures of deconfinement and chiral transitions in the linear sigma model coupled to quarks and the Polyakov loop. We point out the essential role of the causality condition, which requires that any point of the system should rotate slower than the velocity of light. We show that imposing this physical requirement leads to inhibition of the splitting between the chiral and confining transitions, which becomes negligibly small ($\Delta T \sim 1$~MeV or less) for experimentally relevant, slow angular velocities $\Omega \sim 10$~MeV of a $(5-10)$~fm-sized systems. Moreover, the boundedness of the system has a much bigger effect on temperature splitting than the rotation itself: the splitting reaches 10~MeV in a small, one-fermi-sized non-rotating system. The temperature splitting may, however, become enhanced in an academic limit of ultra-relativistic regimes when the boundary of the system rotates at near-to-light velocities.
\end{abstract}

\maketitle

\section{Introduction}

The experimental observation of highly vortical quark-gluon plasmas created in non-central heavy-ion collisions has sparked significant theoretical interest in elucidating the properties of this unprecedentedly vortical fluid~\cite{STAR:2007ccu, STAR:2017ckg}. The vortical effects of the plasma have been examined through hydrodynamic and transport-based models~\cite{Huang:2020dtn, Becattini:2021lfq}, as well as lattice first-principle approaches rooted in numerical Monte Carlo techniques~\cite{Yamamoto:2013zwa, Braguta:2020biu, Braguta:2021jgn, Braguta:2022str, Braguta:2023kwl, Braguta:2023yjn, Yang:2023vsw, Braguta:2023iyx}. 

The strong, non-perturbative nature of quark-gluon plasma precludes the reliable theoretical exploration of its dynamics using conventional, perturbative analytical methods. This constraint necessitates reliance on effective theoretical models that describe the infrared properties of QCD. On the other hand, numerical techniques provide us with information on the behavior of thermodynamic characteristics, such as critical temperature, often without pointing out a physical mechanism that lies behind the numerical data~\cite{Braguta:2020biu, Braguta:2021jgn, Braguta:2022str, Yang:2023vsw} (see, however, the conjecture of the negative Barnett effect put forward in Ref.~\cite{Braguta:2023tqz}). Although numerical methods often do not directly elucidate the mechanisms underlying non-perturbative effects, the data they generate provide us with invaluable information for constraining effective theoretical models and thereby enhancing our understanding of the underlying physical phenomena. In our paper, we will use one of such models, which is the linear sigma model with quarks~\cite{GellMann1960} improved with a coupling to the Polyakov line (PLSM${}_q$)~\cite{Dumitru:2000in, Scavenius:2001pa, Scavenius:2002ru, Megias:2004hj, Schaefer:2007pw}.

The lattice simulations have revealed that vorticity influences the thermodynamic properties and phase structure of quark-gluon plasma in a somewhat unexpected way, as the critical temperature of the deconfining phase transition appears to be a rising function of the angular velocity both for pure gauge theory~\cite{Braguta:2020biu, Braguta:2021jgn} and for QCD~\cite{Braguta:2022str, Yang:2023vsw}. On the contrary, most theoretical models, including field theoretical approaches and holographic techniques, predict that the critical temperature should be a diminishing function of the angular velocity~\cite{Chen:2015hfc, Jiang:2016wvv, Chernodub:2017ref,  Chernodub:2017ref, Wang:2018sur, Chen:2020ath, Mehr:2022tfq,  Zhao:2022uxc, Chen:2023cjt, Sun:2023yux}. This property is readily understood at the level of quarks: the coupling of the intrinsic angular momentum of the constituents of the system to the angular velocity of its global rotation should break the chiral condensate, thus leaving less condensate to be evaporated by the thermal fluctuations. The breaking of the condensate by rotation is a quark analog of the Barnett effect, which tends to polarize spins and orbital momenta in the direction of the angular velocity~\cite{Barnett:1915uqc}.

The situation on the theoretical side can be improved by phenomenologically tuning the parameters of the system in such a way that the parameters of the original model become dependent on the angular velocity itself~\cite{Sun:2024anu}. However, similarly to the first-principle numerical approaches, the fine-tuned effective infrared models offer somewhat limited help in elucidating the physical mechanism that lies behind this fine-tuning. On the other hand, the inclusion of couplings that depend on the external parameters (temperature, rotation) aligns well with the logic of the Polyakov-loop improved models, in which the Lagrangian that incorporates the dynamics of the loop contains already temperature-dependent couplings~\cite{Fukushima:2003fw, Roessner:2006xn, Farias:2016gmy}. 

One should also comment on the potential risk of double counting of the gluonic degrees of freedom when incorporating the Polyakov loop terms into effective quark models, such as the Nambu–Jona-Lasinio~\cite{Nambu:1961tp, Nambu:1961fr} or quark-meson sigma models. For example, in the linear sigma model with quarks~\cite{GellMann1960}, formulated at zero temperature, the terms that provide the interaction of mesons with quarks and the self-interaction of the mesons are interpreted as a result of the dynamics of gluons. In other words, in this and similar models, the gluon degrees of freedom are already considered to be integrated out. Therefore, an additional interaction of the quark fields with the gluon field corresponding to the Polyakov loop may be regarded as a redundancy. However, one can justify the presence of these terms by arguing that the original quark-meson model~\cite{GellMann1960} has been developed to describe zero-temperature physics, which does not account for the global center symmetry ${\mathbb Z}_3$. The breaking of this symmetry is associated with the finite-temperature deconfinement transition. The information about this transition is carried by the Polyakov loop, which appears only in a finite-temperature system. Moreover, the time-like gauge field should be interpreted as a non-trivial classical background, capturing essential features of the confinement-deconfinement transition in QCD. In contrast, the residual gluon fluctuations, which are integrated out in this and similar models, are accounted for through their contribution to the effective four-fermion interaction terms or the interaction with colorless bound states, such as light mesons. Therefore, the inclusion of the Polyakov loop in the finite-temperature effective model seems to be a legitimate extension of the zero-temperature version of the model \cite{Fukushima:2017csk}.

In our paper, we consider the traditional sigma model in which the model parameters and couplings do not depend on external parameters.

The subject of our paper is devoted to the suggestion that the rotation can split the deconfinement and chiral transitions of QCD put forward in Ref.~\cite{Sun:2023kuu}. This hypothesis sounds very reasonable since the global rotation can affect gluons and quarks in different ways, thus forcing the transition to split into the transitions associated, separately, with gluons (the deconfinement transition) and quarks (the chiral transition). On the other hand, Ref.~\cite{Sun:2023kuu} did not consider the boundary effect of the system. The absence of a transverse boundary incurs, in a rigidly rotating case, superluminal velocities of rotating matter that may lead to the appearance of unphysical artifacts in calculations~\cite{Davies:1996ks}. One could argue, however, that if the superluminal effect appears at a large enough distance from the center or rotation --as it happens for a slowly rotating system-- then the breaking of the causality is not a physically relevant property due to the existence of the finite correlation length in the system. Therefore, the effect of the boundaries on the splitting of the transitions needs to be clarified. 

Unfortunately, the PLSM${}_q$ cannot describe the chiral (or chiral and deconfinement) phase transitions even qualitatively. However, this model, similarly to the one considered in Ref.~\cite{Sun:2023kuu}, serves as a very successful effective model for characterizing the infrared phenomenology of the non-rotating quark-gluon matter. In this paper, we use it to probe the effects associated with the spatial boundedness of the system in the presence of rotation and confront them with the results in the unbounded case~\cite{Sun:2023kuu}. Moreover, we expect, given our results below, that the modification of the effective infrared model by introducing environment-dependent couplings will not modify the qualitative conclusions of our paper. 

The structure of the paper is as follows. In Section~\ref{sec_model}, we describe the formulation of the PLSM${}_q$ model in a non-rotating system. Section~\ref{sec_rotating} is devoted to the description of the rigidly rotating environment. To respect the causality, the rigidly rotating system requires the introduction of the boundaries in the transverse plane, which is perpendicular to the angular velocity vector. The cylindrical boundary quantizes the transverse radial modes, leading to a complication of the energy spectrum and requiring extensive numerical computations. In the same section, we show explicitly how quantization emerges in the case of spectral boundary conditions. Section~\ref{sec_split} investigates the phase diagram, the nature of the phase transitions, and the splitting in chiral and deconfining transitions. The effects of the finite radius of the cylinder, the chemical potential, and, finally, the rotation are studied in detail. The last section is devoted to our conclusions. 

\section{The model}
\label{sec_model}

\subsection{Lagrangian}

The Polyakov-loop enhanced linear sigma model coupled to quarks (PLSM${}_q$) has three types of variables: the quark spinor $\psi(x)$ accounting for $N_f = 2$ flavours and $N_c = 3$ colours, the $O(4)$ chiral fields $\phi =(\sigma,\vec{\pi})$ with the pion isotriplet $\vec{\pi} = (\pi^{+},\pi^{-},\pi^{0})$, and the Polyakov loop variable $L(x)$. They enter the Lagrangian of the model via the following three terms that are associated with the mentioned variables:
\begin{align}
    {\mathcal L} = {\mathcal L}_\phi (\sigma, \vec\pi) + {\mathcal L}_L(L) + {\mathcal L}_q(\psi,\phi,L)\,.
    \label{eq_L_full}
\end{align}
In this section, we will briefly describe the properties of the model in the unbounded case.

\subsubsection{Chiral sector}

The chiral field $\phi = (\sigma, \vec\pi)$ encodes, as it follows from its name, the chiral features of the model, with $\sigma$ playing the role of an approximate order parameter of the chiral transition. The field $\sigma$ tends to zero in the large-temperature, chirally restored phase and is non-vanishing in the low-temperature, chirally broken phase. In QCD with realistic masses of quarks, the chiral transition between the two phases is a smooth crossover, implying the absence of a well-defined thermodynamic transition at finite temperatures. 

The chiral part of the PLSM${}_q$ Lagrangian is given by the first term of Eq.~\eqref{eq_L_full}:
\begin{align}
    {\mathcal L}_\phi(\sigma,\vec\pi) = \frac{1}{2} \bigl(\partial _{\mu}\sigma \partial^{\mu}\sigma + \partial_{\mu} {\vec \pi} \partial^{\mu} {\vec \pi} \bigr) - V_\phi(\sigma ,\vec{\pi})\,,
    \label{eq_L_phi}
\end{align}
where the phenomenological potential of the chiral fields,
\begin{align}
    V_\phi(\sigma ,\vec{\pi}) = & \, \frac{\lambda}{4}(\sigma^{2}+\vec{\pi}^{2} - {\it v}^2)^2 - h \sigma 
    \label{eq_V_phi}\,,
\end{align}
exhibits both spontaneous and explicit breaking of chiral symmetry. The potential gives a non-zero expectation value $\langle \sigma \rangle = f_\pi$ for the $\sigma$ field and a vanishing pion expectation value, $\langle \vec{\pi} \rangle = 0$, providing the masses $m_\sigma$ and $m_\pi$ to the sigma and the pion mesons. Writing $\sigma = \langle \sigma \rangle + \delta \sigma$ and $\vec \pi = \langle \vec \pi \rangle + \delta \vec \pi$, these masses can be obtained from the expansion in terms of small fluctuations $\delta \sigma$ and $\delta \pi$:
\begin{equation}
 V = V_0 + \frac{1}{2} m_\sigma^2 \delta \sigma^2 + \frac{1}{2} m_\pi^2 \delta {\vec \pi}^2 + \dots\,,
 \label{eq_V_phi_fluct}
\end{equation}
with $V_0 = V(\langle \sigma \rangle, \langle \vec{\pi} \rangle) = \frac{\lambda}{4}(\langle \sigma \rangle^2 - v^2)^2 - h \langle \sigma \rangle$ and
\begin{equation}
 m_\sigma^2 = \lambda(3 \langle \sigma \rangle^2 - v^2), \quad 
 m_\pi^2 = \langle \sigma \rangle^2 - v^2,
 \label{eq:masses}
\end{equation}
where we imposed $\langle\vec{\pi}\rangle = 0$ as it follows from the minimum of the potential~\eqref{eq_V_phi}.

The analysis of the linear sigma model is usually done in a mean-field approach, in which the ``slow'' meson sector is considered a classical field. In contrast, quarks are treated as quantum fields that represent ``fast'' degrees of freedom. The parameters of the Lagrangian \eqref{eq_L_phi} are chosen to match the low-energy phenomenology~\cite{Scavenius:2000qd}. The masses in Eq.~\eqref{eq:masses} serve as the model parameters that coincide with corresponding meson masses in the vacuum, i.e., at zero temperature, $T=0$, and in the absence of rotation, $\Omega = 0$: $m_\pi = 138\,\mbox{MeV}$ and $m_\sigma = 600\,\mbox{MeV}$. In the absence of electromagnetic interactions, the masses of the charged, $\pi^\pm$, and neutral, $\pi^0$, mesons do not split. 

The explicit symmetry-breaking term is determined by the PCAC relation, $h = f_{\pi} m_{\pi}^2$. Consequently, in the vacuum state, $v^2 = f^2_{\pi} - m^{2}_{\pi} / \lambda$ and $m^2_{\sigma} = 2 \lambda f^2_{\pi} + m^2_{\pi}$. By selecting the mass of the $\sigma$ meson $m_\sigma = 600$ MeV, one gets a reasonable value for the quartic interaction coupling, $\lambda = 20$. If the explicit symmetry breaking term were absent, $h = 0$, then the model~\eqref{eq_L_phi} would experience a second-order phase transition~\cite{Pisarski:1983ms} from the chirally broken symmetry phase to the chirally restored phase at the critical temperature $T_c = \sqrt{2} v$. The explicit symmetry-breaking term removes the exact chiral $O(4)$ symmetry of this theory and alters the phase transition into a smooth crossover in consistency with the QCD phenomenology. Notice that the pion fluctuations play no major role in the chiral phase transition~\cite{Scavenius:2000bb} for the case of the chiral transition (see also a discussion in Ref.~\cite{Skokov:2010wb}). Therefore, below, we focus only on the sigma direction of the chiral sector.

\subsubsection{Confining sector}

The confining properties of the system are accounted for by the second term in Eq.~\eqref{eq_L_full}, which describes the complex-valued Polyakov loop variable,
\begin{align}
L(x) & = \frac{1}{3} {\rm Tr}\, \Phi(x)\,, \nonumber\\ 
\Phi & = {\mathcal P} \exp \Bigl[-i \int\limits_0^{1/T}
{\rm d}\, x_4 \mathcal{A}_4(\vec x, x_4) \Bigr]\,.
\label{eq_L}
\end{align}
The integration takes place along a closed loop in the compactified imaginary time $\tau \equiv x_4$, where $\mathcal{A}_4 = i \mathcal{A}_0$ is the matrix-valued temporal component of the Euclidean gauge field $\mathcal{A}_\mu$ and the symbol ${\mathcal P}$ denotes path ordering that insures the gauge-covariance of the whole expression~\eqref{eq_L}. In the PLSM${}_q$, the Polyakov loop $L$ plays the role of a homogeneous classical scalar field, which can be associated with a coordinate-independent gluon field $\mathcal{A}_4 = \frac{1}{2}(t_3\,  {\mathcal A}_4^{(3)} + t_8\, {\mathcal A}_4^{(8)})$ as $\Phi = \exp(-i \mathcal{A}_4/T)$, where $t_3$ and $t_8$ are the diagonal generators of the $SU(3)$ gauge group.

The Polyakov loop serves as an approximate order parameter for the confining properties of QCD. The expectation value of the Polyakov loop~\eqref{eq_L} vanishes at zero temperature, signaling the absence of the free quarks at $T=0$ and taking a finite value at the high temperature in the quark-gluon plasma phase. The confining properties of the two phases are also smoothly connected. In a pure Yang-Mills theory (in the limit of QCD when the masses of quarks become infinitely large), the Polyakov loop becomes an exact order parameter that vanishes (is a finite quantity) in the whole low-temperature (high-temperature) phase. The Lagrangian of the theory is invariant under the center ${\mathbb Z}_3$ symmetry, $L \to e^{2 \pi n i/3} L $, with $ n=0,1,2$, which is pertinent to the pure gauge theory. At the same time, a coupling to dynamical quarks breaks the ${\mathbb Z}_3$ symmetry explicitly.

The Polyakov loop is coupled to the rest of the fields in the linear sigma model, similar to the case of the Nambu--Jona-Lasinio model as discussed in Refs.~\cite{Fukushima:2003fw, Roessner:2006xn, Roessner:2007gha}. The second term in Eq.~\eqref{eq_L_full},
\begin{align}
    {\mathcal L}_L = - V_L(L,T)\,,
    \label{eq_L_L}
\end{align}
is the potential term of the Polyakov action in the pure Yang-Mills theory not coupled to quarks. We treat the Polyakov loop in a mean-field approach, disregarding a kinetic term for this variable.

We use a specific form for the phenomenological potential of the Polyakov loop~\cite{Roessner:2007gha}:
\begin{align}
    & \frac{V_L(L,T)}{T^4} = -\frac{1}{2}a(T)\,L^*L 
    \label{eq_V_L} \\
    & \quad + b(T)\,\ln\left[1-6\,L^*L+4\left({L^*}^3+L^3\right) - 3\left(L^*L\right)^2\right]\,,
\nonumber
\end{align}
with the parameters
\begin{subequations}
\begin{align}
a(T) = & \, a_0 + a_1\left(\frac{T_0}{T}\right) +a_2\left(\frac{T_0}{T}\right)^2\,,\\ 
b(T) = & \, b_3\left(\frac{T_0}{T} \right)^3\,,
\end{align}
\label{ew_ab_T}
\end{subequations}\!\!\!%
where $T_0$ is the critical temperature in the pure gauge case: $T_0 \equiv T_{SU(3)} = 270\, \mbox{MeV}$. The phenomenological parameters in Eq.~\eqref{ew_ab_T} are
\begin{align}
\begin{array}{llllll}
a_0 & = & 16\,\pi^2/45 \approx 3.51\,, & \qquad a_1 & = & -2.47\,, \\[1mm]
a_2 & = & 15.2\,,                      & \qquad b_3 & = & -1.75\,.
\end{array}
\label{eq_V_parameters}
\end{align}

The set of parameters~\eqref{eq_V_parameters} satisfies, with reasonable accuracy, the following 
requirements emergent from thermodynamics of $SU(3)$ Yang-Mills theory~\cite{Roessner:2007gha}: (i) the Stefan-Boltzmann limit is reached at $T\to \infty$; (ii) a first-order phase transition takes place at $T=T_0$; (iii) the potential describes the existing lattice data for the Polyakov loop and the thermodynamic functions such as pressure, energy density, and entropy.

The definition~\eqref{eq_L} of the Polyakov loop implies that the value of this quantity is limited to the interval $|L^*| \leqslant 1$ and $|L| \leqslant 1$, which is also consistent with a logarithmic divergence of the potential~\eqref{eq_V_L}. According to the parameter set~\eqref{eq_V_parameters}, the value $L^* = L = 1$ can only be reached in the limit $T \rightarrow \infty$. In the confined phase, $T < T_0$, the potential has one trivial minimum, $L = 0$. As the temperature rises and the system undergoes the phase transition, the single minimum at $L = 0$ of the potential splits into three degenerate minima labeled by the ${\mathbb Z}_3$ variable. Thus, in the deconfined phase, the ${\mathbb Z}_3$ symmetry is spontaneously broken, and $\langle L \rangle \neq 0$.

One can argue that in the presence of vorticity, the rotation effects should be incorporated into the Polyakov loop potential because this potential describes the (de)confinement phenomenon in SU(3) gauge theory while, at the same time, the deconfinement transition is affected by the rotation of the same pure gauge matter~\cite{Braguta:2020biu, Braguta:2021jgn}. This modification of the potential is related to the very fundamental non-perturbative mechanism of confinement and, therefore, cannot be taken into account analytically.\footnote{For a perturbative calculation of a rotational contribution to an effective potential of the Polyakov loop, see Refs.~\cite{Chen:2022smf, Chen:2024tkr}.} Hence, the rotation-dependent terms in the potential should be introduced in a phenomenological way, as it was done in Ref.~\cite{Sun:2023yux}. By promoting the parameters of the Polyakov loop potential to phenomenological functions of the angular velocity, one finds that the model, indeed, better describes a particular, volume-averaged set of available lattice data for vortical gluon matter.

However, the recent lattice data have demonstrated that the rotation generates inhomogeneities of the Polyakov loop expectation values in the radial direction with respect to the axis of rotation
~\cite{Braguta:2023iyx}. This effect cannot be described by translation-invariant terms in the potential that were incorporated in the study of Ref.~\cite{Sun:2023yux}. Therefore, in our paper, we follow the standard approach in which the potential does not include additional rotation-dependent contributions~\eqref{eq_V_L}. This approach will allow us to test how well the traditional model describes the rotational phenomena in the quark-gluon plasma.

\subsubsection{Quark sector}

The third term of the full Lagrangian~\eqref{eq_L_full},
\begin{align}
    {\mathcal L}_q = \overline{\psi} \left[i {\slashed D} 
    - g (\sigma + i \gamma_5 \vec{\tau} \cdot {\vec \pi} )\right]\psi\,,
    \label{eq_L_psi}
\end{align}
describes the spinor doublet $\psi = (u,d)^T$ of the light quarks which interacts with the meson fields $\sigma$ and ${\vec \pi}$. The spinor field $\psi$ also interacts with the $SU(3)$ gauge field $A_\mu$ via the covariant derivative:
\begin{align}
    {\slashed D} = \gamma^{\mu} D_\mu\,, \qquad D_\mu = \partial_\mu - i \mathcal{A}_\mu\,.
    \label{eq_D}
\end{align}
The gauge field represents, according to Eq.~\eqref{eq_L}, a nontrivial background due to the Polyakov loop $L$. 

The spontaneous symmetry breaking in the chiral field sector, $\langle\sigma\rangle \neq 0$, also gives the constituent mass $m_q \equiv g \langle\sigma\rangle$ to the quark field. Setting the vacuum mass for the constituent quark to $307$~MeV gives us the Yukawa coupling $g \simeq 3.3$. At low temperatures, quarks are not excited, and the model~\eqref{eq_L_phi} becomes the standard linear $\sigma$-model without quarks~\cite{kapusta_gale_2006}.

Thus, the fermion field $\psi$ couples two other variables: the Polyakov loop $L$ and the chiral field $\phi$, linking confining and chiral properties together. Moreover, the dynamical quark fields $\psi$ are affected by the global rotation. As the quarks interact with the Polyakov loop $L$ and the chiral field $\phi$, the rotation influences both the chiral dynamics and the confining properties of the model. Therefore, this model allows us to study how rotating simultaneously affects color confinement and chiral symmetry breaking.
%

Summarizing the description of the system, we discussed an effective infrared model that describes the low-energy dynamics of quark degrees of freedom that are also coupled to the confining gluonic degrees of freedom. The latter is represented by the gauge field ${\mathcal A}_4$, which is directly related to the confining order parameter, the Polyakov loop. One could think that in this approach, the additional explicit quark-gluon interactions, relevant at $T \neq 0$, might represent an additional redundancy since the gluonic fields are already incorporated in the $T=0$ parameters of the quark-meson model. However, in the adopted approach, these terms are treated differently: the Polyakov loop in quark-gluon terms accounts for the central symmetry that distinguishes the confining and deconfining phases and represents a non-trivial classical gluon background, while the quark-meson and meson-meson terms originate from the $T=0$ gluon fluctuations that are not treated as classical fields.


\subsection{Thermodynamics of nonrotating system}

Before considering the main subject of our paper, the quark system in rotation, we first comprehensively describe the general principle of how the thermodynamics of the non-rotating plasma is determined in the PLSM${}_q$ model. A more detailed review and careful determination of the thermodynamic properties of the model will be reported in our forthcoming paper~\cite{in_preparation}.  

\subsubsection{Thermodynamic potential}

We adopted the mean-field approximation by replacing $\sigma$ and ${\vec \pi}$ by their mean-field (expectation) values as we discussed above, in the ground state, $\langle\vec\pi\rangle \equiv 0$ in all studied phases. The quantity $\langle \sigma \rangle$ determines the strength of the chiral symmetry breaking and can be nonzero.

The partition function of the PLSM${}_q$ follows directly from its Lagrangian~\eqref{eq_L_full}:
\begin{align}
    \mathcal{Z} = & \,
    \exp\Bigl\{ - \frac{V_{3d}}{T} \Bigl[V_\phi(\sigma, {\vec\pi}) + V_L(L) \Bigr] \Bigr\} 
    \label{eq_part}\\
    & \times \int \mathcal{D}\psi \mathcal{D}\Bar{\psi} \exp\left\{- \int_0^\beta d\tau \int_V d^3x \, {\mathcal L}_q(\psi,\sigma,L)  \right\}\,,
\nonumber
\end{align}
where an irrelevant overall normalization factor has been omitted. Equation~\eqref{eq_part} gives us the thermodynamic potential of the model,
\begin{align}
    F(T) \equiv -\frac{T\ln \mathcal{Z}}{V} = V_\phi(\sigma) + V_L(L) + F_{\psi\Bar{\psi}}(\sigma, L)\,,
    \label{eq_F}
\end{align}
where the meson potential $V_\phi$ and the Polyakov loop potential $V_L$ can be read from Eqs.~\eqref{eq_V_phi} and \eqref{eq_V_L}.

The last term in Eq.~\eqref{eq_F} corresponds to the quark contribution to the thermodynamic potential, which is given by the last multiplier in Eq.~\eqref{eq_part}. The quark part can be split into its vacuum and thermal constituents: $F_{\psi\bar{\psi}} = F^{\rm vac}_{\psi\bar{\psi}} + F^{\beta}_{\psi\bar{\psi}}$, where the vacuum contribution to the thermodynamic potential, 
\begin{equation}
 F^{\rm vac}_{\psi\bar{\psi}} = -2N_c N_f \int \frac{d^3p}{(2\pi)^3} E,
 \label{eq_FQuark_vac}
\end{equation}
is infinite. We, however, follow the arguments provided in Ref.~\cite{Scavenius:2000qd} and ignore below the vacuum quark part by setting $F^{\rm vac}_{\psi\bar{\psi}} = 0$. Since now $F_{\psi\bar{\psi}} = F^{\beta}_{\psi\bar{\psi}}$, we will omit the ``thermal'' $\beta$ superscript and instead identify the fermionic grand potential entirely with its thermal contribution~\cite{Scavenius:2000qd, Ratti:2007jf}:
\begin{equation}
    F_{\psi\Bar{\psi}}= -2N_f T \int \frac{d^3p}{(2\pi)^3}
    \sum_{\varsigma = \pm} {\rm Tr}_{c} \ln \left[1+e^{- \beta (\mathcal{E}_\zeta + i \zeta {\mathcal A}_4)}\right] \,,
    \label{eq_F_psi}
\end{equation}
where $\mathcal{E}_\varsigma = E - \varsigma \mu$ represents the effective energy relative to the Fermi level $E_F = \mu$ for particles ($\varsigma = +$) and $E_F = -\mu$ for antiparticles ($\varsigma = -$), with $\mu$ being the quark chemical potential. The trace ${\rm Tr}_c$ over colour indices can be taken explicitly:
\begin{equation}
 F_{\psi\Bar{\psi}} =-2N_fT \sum_{\varsigma = \pm 1} \int \frac{d^3p}{(2\pi)^3}
    F_\varsigma\,,
    \label{eq_FQuark}
\end{equation}
where we introduced the particle and antiparticle contributions, respectively (see Ref.~\cite{Torres-Rincon:2017zbr} for technical details):
\begin{align}
 F_+ &= \ln\left[1+3Le^{-\beta \mathcal{E}_+}+3L^* e^{-2\beta \mathcal{E}_+}+e^{-3\beta \mathcal{E}_+}\right], \nonumber\\
 F_- &= \ln\left[1 + 3L^*e^{-\beta \mathcal{E}_-} + 3Le^{-2\beta \mathcal{E}_-} + e^{-3\beta \mathcal{E}_-}\right].
 \label{eq_Fpm}
\end{align}

\subsubsection{Saddle point approximation}

We find the mean-field values of the fields $\sigma$ and $L$ at thermodynamic equilibrium by minimizing the total free energy with respect to $\sigma$, $L_+ \equiv L$, and $L_- \equiv L^*$ variables:
\begin{align}
    \frac{\partial F}{\partial \sigma} =
    \frac{\partial F}{\partial L} = \frac{\partial F}{\partial L^*} = 0\,.
    \label{eq_saddle}
\end{align}
These equations can be expressed with the help of the following derivatives:
\begin{align} \hskip -3mm
    f_\pm {=} -\frac{1}{3}  \frac{\partial F_\pm}{\partial (\beta \mathcal{E}_\pm)}, 
    \quad
    f_{\pm,+} {=} \frac{1}{3} \frac{\partial F_\pm}{\partial L_+}, 
    \quad
    f_{\pm,-} {=} \frac{1}{3} \frac{\partial F_\pm}{\partial L_-}, 
\end{align}
From Eq.~\eqref{eq_Fpm}, it is not difficult to see that the above quantities take the following explicit form:
\begin{subequations}
\begin{align}
 f_\pm & =\frac{L_\pm e^{-\beta \mathcal{E}_\pm} + 2 L_\mp e^{-2\beta\mathcal{E}_\pm} + e^{-3\beta \mathcal{E}_\pm}}{1 + 3 L_\pm e^{-\beta \mathcal{E}_\pm} + 3L_\mp e^{-2 \beta \mathcal{E}_\pm} + e^{-3 \beta \mathcal{E}_\pm}}, 
 \label{eq_f}\\
 f_{\pm,\pm} & = \frac{e^{-\beta \mathcal{E}_\pm}}{1 + 3 L_\pm e^{-\beta \mathcal{E}_\pm} + 3L_\mp e^{-2 \beta \mathcal{E}_\pm} + e^{-3 \beta \mathcal{E}_\pm}}, 
 \label{eq_fL}\\
 f_{\pm,\mp} & = \frac{e^{-2\beta \mathcal{E}_\pm}}{1 + 3 L_\pm e^{-\beta \mathcal{E}_\pm} + 3L_\mp e^{-2 \beta \mathcal{E}_\pm} + e^{-3 \beta \mathcal{E}_\pm}}.
 \label{eq_fLb}
\end{align}
\label{eq_fall}
\end{subequations}

One obtains the following explicit form of the saddle-point equations~\eqref{eq_saddle}:
\begin{align}
 \frac{\partial F}{\partial \sigma} & = \frac{\partial V_\phi}{\partial \sigma} + g \langle \Bar{\psi} \psi \rangle\,, 
 \label{eq_dF_ds}\\
 \frac{\partial F}{\partial L_\pm} & = \frac{\partial V_L}{\partial L_\pm} - 6 N_f T \sum_{\varsigma = \pm 1} \int \frac{d^3p}{(2\pi)^3} f_{\varsigma, \pm}\,,
 \label{eq_dF_dL}
\end{align}
where we identified the fermion condensate in the first equation, 
\begin{equation}
 \langle \Bar{\psi} \psi \rangle = 6 N_f g \sigma \sum_{\varsigma = \pm 1} \int \frac{d^3p}{(2\pi)^3 E} f_\varsigma\,,
 \label{eq_ppsi}
\end{equation}
while the derivatives of the mesonic and Polyakov loop potentials read as follows:
\begin{subequations}\label{eq_dVdp}
\begin{align}
 \frac{\partial V_\phi}{\partial \sigma} = & \, \lambda(\sigma^2 - v^2) \sigma - h\,, \label{eq_dVds}\\
 \frac{\partial V_L}{\partial L_\pm}  = & \, - \frac{6 T^4 b(T) (L_\mp - 2 L_\pm^2 + L_\mp^2 L_\pm)}{1 - 6 L^* L + 4 (L^*{}^3 + L^3) - 3(L^* L)^2} \nonumber\\
 & - \frac{T^4}{2} a(T) L_\mp. \label{eq_dVdL}
\end{align}
\end{subequations}

One immediately notices that Eqs.~\eqref{eq_dVdp} contain cubic powers of the order parameters, which implies that the saddle-point equations \eqref{eq_saddle} should provide us with multiple solutions. The physically permitted solutions should satisfy the following constraints:
\begin{equation}
 0 < \sigma \le f_\pi\,, \qquad\ 0 \le L, L^* < 1\,.
 \label{eq_constraints}
\end{equation}
The first condition in Eq.~\eqref{eq_constraints} appears on physical grounds as we expect that both thermal fluctuations and finite-density environment restore the chiral symmetry instead of enhancing its breaking. This assertion is equivalent to the statement that the zero-temperature QCD vacuum, defined as a state at $T = 0$ and $\mu = 0$ (and, appropriately, at $\Omega = 0$), the highest value of the mean-field (expectation) value of the $\sigma$ meson field gives the quarks their maximal (constituent) dynamical mass, $m_q = g \sigma$, in the whole phase diagram. The value of $\sigma$ is a positive quantity because of the presence of the last term in the potential~\eqref{eq_V_phi} on the chiral field $\phi$, with $h > 0$ fixed by the phenomenology.

The second restriction in Eq.~\eqref{eq_constraints} also has two bounds. The lower bound appears due to the property that in the vacuum, the Polyakov loop expectation value vanishes, $L = L^* = 0$, reflecting the fact that the addition of an infinitely heavy (anti-)quark requires an infinitely large change in the thermodynamic potential of the system. Therefore, the corresponding contributions of one quark, $F_Q = - T \ln L$, and of one anti-quark,  $F_{\bar Q} = - T \ln L^*$, are infinite.
Notice that within the scope of the Polyakov loop models, the two conjugate fields $L$ and $L^*$ are treated as real but independent numbers. It means, in particular, that the free energy of a heavy (anti-)quark, defined above, is a real number.

As the medium becomes more energetic (either denser, at higher chemical potential $\mu$; or hotter, at higher temperature $T$), the quark-gluon medium enters a deep deconfinement phase, which is achieved when the Polyakov loop and its conjugate approach their maximal expectation values, $L, L^* \rightarrow 1$. Values outside the intervals indicated in Eq.~\eqref{eq_constraints} cannot represent physically realizable systems. The upper bound on the Polyakov loop~\eqref{eq_constraints} arises due to the mathematical definition of the Polyakov loop~\eqref{eq_L}.

Moreover, if even Eqs.~\eqref{eq_saddle} give us multiple solutions satisfying the physical constraints of Eqs.~\eqref{eq_constraints}, we should select the true ground state distinguishing it from unstable and (meta)stable states. The unstable solutions correspond to points of, respectively, local maxima or saddle points of the total free energy, $F$. The lowest-free-energy minimum represents the true stable ground state of the system. Quantum and thermal fluctuations typically drive a quantum system from metastable states, associated with local free energy minima, to the true ground state, corresponding to its global minimum.

In order to evaluate the free energy~\eqref{eq_F}, we add the mesonic potential~\eqref{eq_V_phi} and the Polyakov-loop potential~\eqref{eq_V_L} to the fermionic contribution~\eqref{eq_F_psi}. The latter expression can be simplified using the integration by parts,
\begin{equation}
 F_{\psi\Bar{\psi}} \equiv - P_{\psi\Bar{\psi}} = - 6N_f \sum_{\varsigma = \pm 1} \int \frac{d^3p}{(2\pi)^3} \frac{p^2}{3E} f_\varsigma\,.
 \label{eq_P}
\end{equation}
In consistency with thermodynamics of fermionic gases, the fermionic pressure~\eqref{eq_P} is given by the expectation value of the canonical expression: $P_{\psi\Bar{\psi}} = -\frac{i}{6} \Bar{\psi} \boldsymbol{\gamma} \cdot \overleftrightarrow{\boldsymbol{\nabla}} \psi$.

\subsection{Phase diagram of non-rotating plasma}

For a static plasma, the thermodynamic limit has a well-defined meaning, which allows us to check our algorithms for determining the phase diagram before proceeding, in the next section, to the more complicated case of the rigidly-rotating plasma. 

In terms of the chiral properties, the phase diagram of the theory contains two phases: the chirally broken phase and the chirally restored phase, separated by a transition point. At low values of the vector chemical potential, the phase diagram is known to possess a smooth crossover, which does not contain any thermodynamic singularity. Therefore, both phases are analytically connected. At higher values of the chemical potential, the division between the phases turns into a first-order phase transition separated from crossover by an end-point at which the system experiences a second-order phase transition. 

A similar behavior is also pertinent to the Polyakov loop expectation value, which is an order parameter for the confining (low-temperature) and deconfining (high-temperature) phases. Below, we demonstrate these properties, highlighting the subtleties of the determination of the critical points in the phase diagram.

\begin{figure}[ht]
    \centering
\includegraphics[width=0.9\linewidth]{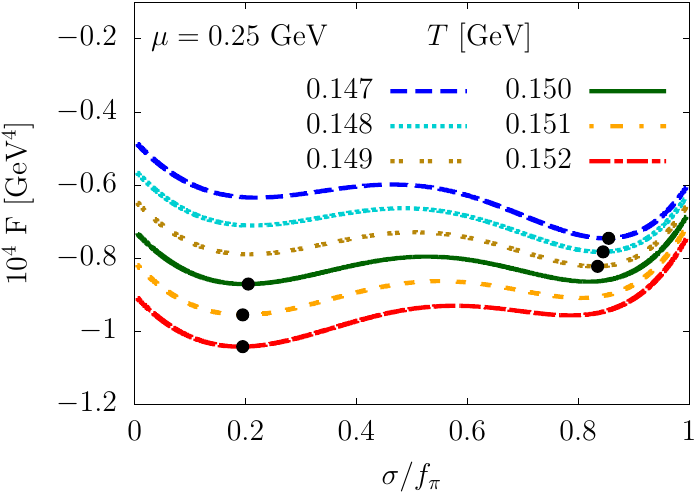}
    \caption{Dependence of the grand canonical potential $F$ in the PLSM${}_q$ model on the expectation value of the $\sigma$ field at the fixed value of the vector chemical potential, $\mu = 0.25$~GeV. Various temperatures near the phase transition point are shown. The small filled circles indicate the value of the normalized expectation value of the sigma field, $\sigma / f_\pi$, which corresponds to the global minimum of $F$ at each temperature $T$.}
    \label{fig_extrema}
\end{figure}

Figure~\ref{fig_extrema} illustrates the typical behavior of the full thermodynamic potential~\eqref{eq_F} -- excluding the vacuum part -- as a function of the $\sigma$ mean field at the vector chemical potential $\mu = 0.25$~GeV, which is sufficiently large to allow for a first-order phase transition. In this figure, we consider the field $\sigma$ as a free (external) parameter and find the values of the $L$ and $L^*$ that minimize $F$ at fixed values of $\sigma$, $T$, and $\mu$. In the vicinity of the phase transition, the free energy $F$ develops multiple extrema corresponding to two minima and one maximum. The thermodynamically favored phase corresponds to the global minimum, while the other minimum represents a metastable state and the maximum gives us an unstable point. 

At this particular value of the vector chemical potential, $\mu = 0.25$~GeV, the phase transition appears at $T_c \simeq 0.15$~GeV, implying that the theory stays in the chirally-broken phase at lower temperatures and in the chirally-restored phase at higher temperatures. Notice that the chiral restoration in the high-temperature phase has an approximate meaning. Indeed, Fig.~\ref{fig_extrema} shows that at the phase transition point, the mean value of the field $\sigma$ drops rather strongly but still does not vanish at the high-temperature phase at $T > T_c$. The same remark applies also to the deconfinement transition.

\begin{figure}[ht]
    \centering
\includegraphics[width=0.95\linewidth]{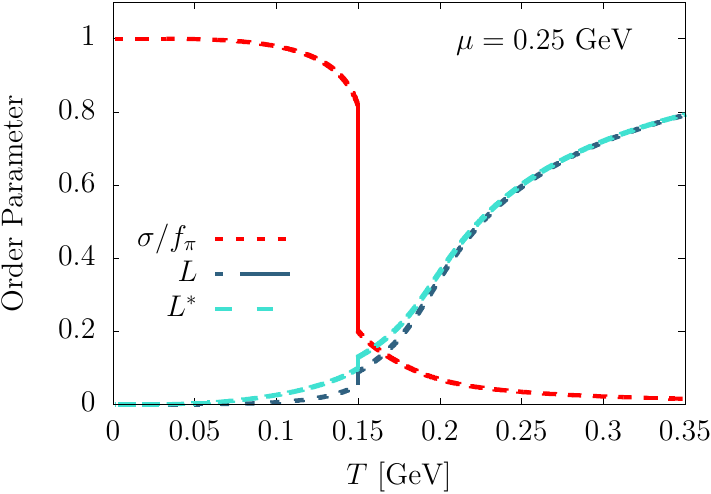}
    \caption{Temperature dependence of the normalized order parameter for the chiral phase transition, $\sigma/f_\pi$, and the order parameters for the confinement-deconfinement transitions, $L$ and $L^*$, at $\mu=0.25$~GeV.}
    \label{fig_pres}
\end{figure}

Figure~\ref{fig_pres} shows the order parameter for the chiral transition, $\sigma$, normalized to its vacuum value $\sigma_{\rm vac} = f_\pi$ as a function of temperature. In the same figure, we also show the expectation value of the Polyakov loop $L$ and its conjugate $L^*$, which are the confinement order parameters. This figure clearly shows that the increase in temperature or chemical potential drives the system from the confinement (chirally broken) to the deconfinement (chirally restored) phase, as expected. 

Notice that the presence of finite density of quark matter, characterized by the non-zero chemical potential, $\mu \neq 0$, leads to the explicit breaking of the charge-conjugation symmetry, such that the properties of quarks become different from the properties of anti-quarks. In particular, the free energy of an infinitely heavy quark, $F_Q = - T \ln L$, and the free energy of an anti-quark, $F_{\bar Q} = - T \ln L^*$, differ from each other (we remind that both $L$ and $L^*$ are real-valued quantities). Figure~\ref{fig_pres} illustrates this behavior: adding another quark to a quark-rich medium (with $\mu>0$) is more costly than adding an anti-quark to the same medium (i.e., $F_{Q} > F_{\bar Q}$ for all values of $\mu > 0$). 

We are now interested in identifying a line in the $(T-\mu)$ plane of parameters where exactly the transition takes place. In the case of a first-order phase transition, which is realized in the presence of a nonzero quark density, it is sufficient to find the point along the temperature direction (that is, at a fixed $\mu$ and varying $T$) where the order parameter(s) exhibit a discontinuity. Typically, the discontinuity in the $\sigma$ condensate is accompanied by discontinuities in the other thermodynamic parameters, including in the values of the Polyakov loops $L$ and $L^*$. While the intuitive picture of the (phase) transition behavior is clearly seen from Fig.~\ref{fig_pres} at a high $\mu$, it is not the case at lower densities, where the system undergoes a so-called crossover transition without encountering a thermodynamic singularity. 

In the absence of a discontinuity at low $\mu$'s, the system goes continuously from one phase to the other as the thermodynamic parameters change. In practical terms, discriminating between a first-order transition and a sharp crossover transition poses a formidable numerical challenge. Pinpointing the exact position of the pseudo-critical parameters of the crossover transition has a certain degree of arbitrariness, which depends on the choice of the observables used to identify the crossover. 

\begin{figure}[ht]
    \centering
\includegraphics[width=0.95\linewidth]{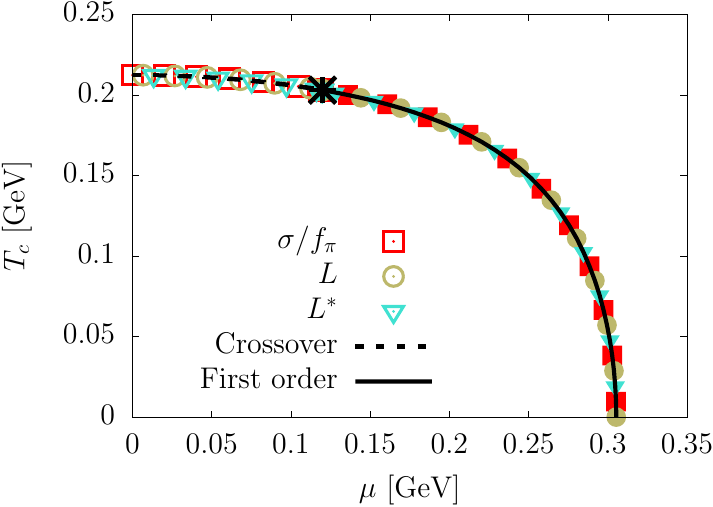}
    \caption{Phase diagram for unbounded static $\mathrm{PLSM}_q$ model in $T - \mu$ plane for the normalized order parameter of the chiral transition $\sigma/f_\pi$, as well as the confining transition, $L$ and $L^*$. The phase diagram features a crossover transition line at lower vector chemical potentials $\mu$, which is marked by the dashed line. The first-order phase transition at higher values of $\mu$ is shown by the solid line. The critical end-point that separates these regimes occurs at $(T_c, \mu_c)= (0.2043, 0.1123)$~GeV. It is marked by the star ``*''.}
    \label{fig_pd}
\end{figure}

In our work, the point of the crossover is identified as an inflection point with respect to the relevant order parameters $\sigma$, $L$, and $L^*$. The inflection is searched along a straight line that is connected with the origin of the parameter space, $\mu = T = 0$. The resulting phase diagram is shown in Fig.~\ref{fig_pd}. Qualitatively, it agrees with known results obtained in the thermodynamic limit~\cite{Schaefer:2007pw}.\footnote{The predictions of the effective models may differ quantitatively depending on number of quark flavors, values of the model couplings, treatment of field fluctuations and implementation of renormalization (see, for example, Ref.~\cite{Kashiwa:2007hw}).} There is no splitting between the deconfining and chirality-restoring transitions neither for the thermodynamic phase transition at higher $\mu$ nor for the crossover regime at the lower-density part of the phase diagram at smaller values of the chemical potential~$\mu$. As we will see below, this property will be lost for a system where two of the three spatial directions are bounded, as it is pertinent to the rotating systems. 

At the low-$T$ high-$\mu$ corner of the phase diagram of Fig.~\ref{fig_pd}, a strong chiral transition, which is marked by a substantial discontinuity in the chiral condensate, is accompanied by very modest discontinuities in the Polyakov loop $L$ as well as its conjugate $L^*$. Therefore, the free energies of a heavy quark and a heavy anti-quark experience a relatively insignificant change across the transition. This observation provokes the natural question of whether this transition can be interpreted as a deconfining transition. This question is more than legitimate as the Polyakov loop, strictly speaking, does not play the role of an order parameter in the presence of dynamical quarks that break the center ${\mathbb Z}_3$ symmetry explicitly. Physically, it means that even at vanishing chemical potential, the dynamical quark fields screen the chromoelectric field of a heavy test quark $Q$. The process proceeds via the breaking of the confining string attached to the heavy quark through the $q \bar q$ pair creation. Therefore, the free energy associated with the Polyakov loop $L$ corresponds to the free energy of the $Q \bar q$ heavy-light meson, which takes a finite value even in the hadronic phase. Consequently, the expectation value of the Polyakov loop, strictly speaking, does not vanish in the hadron phase~\cite{Greensite:2003bk}.

This effect becomes even more pronounced in the presence of a finite chemical potential, as the Polyakov loop gets screened by the on-shell quarks that are already present in the system. Moreover, when the density of quarks is high, the mean inter-quark density is smaller than the distance at which the QCD string gets formed. Therefore, the confining property is, in a rigorous sense, lost as the confining string of the heavy quark cannot develop in the system. This analogy with QCD has a qualitative nature since the effective PLSM${}_q$ model does not possess the dynamical gluon degrees of freedom responsible for the string formation. Still, using continuity arguments that connect analytically the low-$T$ high-$\mu$ corner of the phase diagram with the high-temperature plasma, we interpret the weak discontinuity in the Polyakov loop as a signature of a deconfining transition. It is necessary to mention that at the high quark density, the phase structure of QCD is expected to become more complex as the ground state may be described by more involved quark condensates~\cite{Fukushima:2013rx}. We do not expand further our speculations about the high-density and low-temperature corner of the QCD phase diagram as this parameter goes beyond the applicability of the simple low-energy models such as PLSM${}_q$.

\section{The PLSM${}_q$ model in rigid rotation}
\label{sec_rotating}

Below, we discuss the formulation of the PLSM${}_q$ system at finite angular momentum. To facilitate the analysis of the phase diagram, we will employ corotating coordinates, defined by the angular velocity $\boldsymbol \Omega$ entering the grand canonical ensemble. To avoid superluminal corotating velocities, the system must be enclosed in a cylindrical volume, where suitable boundary conditions must be enforced in the plane transverse to the angular velocity vector. 

\subsection{Co-rotating frame}

The description of the fermionic fields in the co-rotating reference frame follows the standard procedure~\cite{Ambrus:2014uqa, Chen:2015hfc, Chernodub:2016kxh}. We switch to corotating coordinates, defined with respect to the static cylindrical coordinates of the inertial laboratory reference frame (denoted by subscript ``lab'') by 
\begin{equation}
 t = t_{\rm lab}\,, \qquad \varphi = \varphi_{\rm lab} - \Omega t_{\rm lab}\,,
\end{equation}
where $\Omega \equiv |{\boldsymbol{\Omega}}| > 0$ represents the angular frequency of the corotating observer. The line element $ds^2 = g_{\mu\nu} dx^\mu dx^\nu$ in the corotating coordinates reads as follows:
\begin{equation}
 ds^2 = (1 - \rho^2 \Omega^2) dt^2 - 2\rho^2 \Omega dt d\varphi - d\rho^2 - \rho^2 d\varphi^2 - dz^2\,.
\end{equation}
According to the standard procedure, the curvilinear metric $g_{\mu\nu}$ can be brought to Minkowski form by introducing the non-holonomic, orthonormal tetrad (vierbein) vector field, $e_{\hat{\alpha}} = e_{\hat{\alpha}}^\mu \partial_\mu$, and the associated one-forms $\omega^{\hat{\alpha}} = \omega^{\hat{\alpha}}_\mu dx^\mu$, defined by 
\begin{align}
 e_{\hat{t}} &= \partial_t + \Omega(y \partial_x - x \partial_y)\,,  \qquad
 & \omega^{\hat{t}} = dt\,,  \nonumber\\
 e_{\hat{x}} &= \partial_x\,, \qquad  
 & \omega^{\hat{x}} = dx - \Omega y dt\,,
 \label{eq_ew}\\
 e_{\hat{y}} &= \partial_y\,, \qquad 
 & \omega^{\hat{y}} = dy + \Omega x dt\,,  \nonumber \\
 e_{\hat{z}} &= \partial_z, \qquad 
 & \omega^{\hat{z}} = dz\,, \nonumber
\end{align}
such that $g_{\mu\nu} = \eta_{\hat{\alpha}\hat{\beta}} \omega^{\hat{\alpha}}_\mu \omega^{\hat{\beta}}_\nu$ and $\omega^{\hat{\alpha}}_\mu e^\mu_{\hat{\beta}} = \delta^{\hat{\alpha}}_{\hat{\beta}}$. 

The Cartan coefficients associated with the tetrad are defined by the commutation relation $[e_{\hat{\alpha}}, e_{\hat{\rho}}] = c_{\hat{\alpha}\hat{\rho}}{}^{\hat{\sigma}} e_{\hat{\sigma}}$, where $e_{\hat{\alpha}}$ are considered as differential operators~\eqref{eq_ew}.  Taking into account the following explicit form of the nontrivial pairs of these commutation relations,
\begin{equation}
 [e_{\hat{t}}, e_{\hat{x}}] = \Omega e_{\hat{y}}, \qquad 
 [e_{\hat{t}}, e_{\hat{y}}] = -\Omega e_{\hat{x}},
\end{equation}
we find the following non-vanishing Cartan coefficients:
\begin{equation}
 c_{\hat{t}\hat{x}}{}^{\hat{y}} = c_{\hat{y}\hat{t}}{}^{\hat{x}} = \Omega\,.
\end{equation}
The connection coefficients $\Gamma_{\hat{\rho}\hat{\sigma} \hat{\alpha}} = \frac{1}{2} (c_{\hat{\rho}\hat{\sigma} \hat{\alpha}} + c_{\hat{\rho}\hat{\alpha} \hat{\sigma}} - c_{\hat{\sigma} \hat{\alpha} \hat{\rho}})$ vanish identically, except for the following component:
\begin{equation}
 \Gamma_{\hat{x}\hat{y}\hat{t}} = \Omega\,.
\end{equation}

The Dirac action can be written in curvilinear coordinates as
\begin{align}
 S & = \int d^4x \sqrt{-g} \mathcal{L}_D\,, \nonumber \\ 
 \mathcal{L}_D & = \overline{\psi} \left[i \slashed{D} + \slashed{\mathcal A} + \mu \gamma^0  - g \sigma \right] \psi,
\end{align}
where $D_\mu = \partial_\mu + \Gamma_\mu$ is the covariant derivative and $\slashed{\mathcal A} = \gamma^\mu {\mathcal A}_\mu$ is the Abelian gauge field, corresponding to the Cartan subgroup of the SU(3) gauge group, written in the Feynman slash notation. Here, we have also included the vector chemical potential $\mu$ to account for the shifts of the Dirac sea in the presence of matter--anti-matter imbalance. The spin connection coefficient is given by $\Gamma_\mu = -\frac{i}{2} \omega_\mu^{\hat{\alpha}} \Gamma_{\hat{\beta}\hat{\gamma}\hat{\alpha}} S^{\hat{\beta}\hat{\gamma}}$, where the structure $S^{\hat{\beta}\hat{\gamma}} = \frac{i}{4} [\gamma^{\hat{\beta}}, \gamma^{\hat{\gamma}}]$ represent the spin part of the generators of the Lorentz transformations. 

We choose the gamma matrices with respect to the tetrad vector fields in the Dirac representation,
\begin{equation}
 \gamma^{\hat{t}} = \begin{pmatrix}
  1 & 0 \\ 0 & -1
 \end{pmatrix}, \quad 
 \gamma^{\hat{i}} = \begin{pmatrix}
  0 & \sigma^{\hat{i}} \\ -\sigma^{\hat{i}} & 0
 \end{pmatrix},
\end{equation}
with $\sigma^{\hat{i}}$ being the Pauli spin matrices:
\begin{equation}
 \sigma^{\hat{x}} = \begin{pmatrix}
  0 & 1 \\ 1 & 0
 \end{pmatrix}, \quad 
 \sigma^{\hat{y}} = \begin{pmatrix}
  0 & -i \\ i & 0
 \end{pmatrix}, \quad 
 \sigma^{\hat{z}} = \begin{pmatrix}
  1 & 0 \\ 0 & -1
 \end{pmatrix}.
\end{equation}
With the above convention, the single non-vanishing spin connection coefficient takes the simple form, $\Gamma_{\hat{t}} = -i \Omega S^z$, implying
\begin{equation}
 iD_{\hat{t}} = i\partial_t + \Omega J^z,  
\end{equation}
where
\begin{align}
 J^z = -i \partial_\varphi + S^z\,, 
 \qquad
 S^z  = \frac{1}{2} \begin{pmatrix} \sigma^z & 0 \\ 0 & \sigma^z \end{pmatrix},
\end{align}
are the operator of the total angular momentum and its spin part, respectively. 

Then, the Dirac equation for the spinor $\psi$ in the co-rotating reference frame reads as follows:
\begin{align}
 [\gamma^0 (i \partial_t + \mathcal{A}_0 + \mu + \Omega J^z) + i \boldsymbol{\gamma} \cdot \boldsymbol{\nabla} - g\sigma] \psi &= 0\,.
 \label{eq_rot_Dirac_eq}
\end{align}
Its charge conjugate that satisfies the same equation is 
\begin{align}
    \psi^c = i \gamma^2 \psi^*{\Bigl|}_{{\mathcal A}^\mu \rightarrow -{\mathcal A}^\mu, \, \mu \rightarrow -\mu}\,.
\end{align}

The Dirac equation in the rotating reference frame~\eqref{eq_rot_Dirac_eq} implies a simple linear relation,
\begin{align}
    H = H_{\rm lab} - {\boldsymbol{\Omega}} \cdot {\boldsymbol{J}}\,,
    \label{eq_H_Hlab}
\end{align}
between the Dirac Hamiltonian in the rotating reference frame, $H = i \partial_t$, and the Hamiltonian in the laboratory reference frame, $H_{\rm lab} = i \partial_{t_{\rm lab}}$. The change in the Hamiltonian of the system~\eqref{eq_H_Hlab} as we move from one reference system to the other is given by the coupling of the angular velocity $\boldsymbol{\Omega}$ and the total angular momentum $\boldsymbol{J}$. The same property applies also to classical mechanical systems and to thermodynamics~\cite{LL1, LL5}. 

Below, the thermodynamic quantities will be evaluated in the Euclidean spacetime in the Matsubara formalism. Under the Wick rotation to the imaginary time, $ t \to \tau = i t$, the partition function of the model takes the form: 
\begin{align}
 Z & = \int [i d\psi^\dagger][d\psi] 
 \exp(S_E)\,, \nonumber \\ 
 S_E & = \int_0^\beta d\tau \int_V d^3x \, \mathcal{L}_E\,,
 \label{eq_cyl_SE}
\end{align}
where $V$ represents the system volume (a cylinder of radius $R$ and infinite longitudinal extent). The Euclidean Lagrangian $\mathcal{L}_E$ reads as follows:
\begin{equation}
 \mathcal{L}_E = \overline{\psi} \left[\gamma^0(-\partial_\tau + \Omega J^z + i\mathcal{A}_4 + \mu) + i \boldsymbol{\gamma} \cdot \boldsymbol{\nabla} - g \sigma\right) \psi\,.
 \label{eq_LE}
\end{equation}

\subsection{Dirac modes and spectral boundaries}

In order to understand the thermodynamics of the rotating system, we need to calculate the eigenenergy of the Hamiltonian $H = i \partial_t$ in the rotating reference frame, which is bound in the transverse directions by a cylindrical boundary. The eigenvalues are characterized by the quantum numbers associated with the angular momentum $J^z$, the longitudinal momentum $P^z = -i \partial_z$, and the helicity $h = \mathbf{S} \cdot \mathbf{P} / p$ operators. These operators commute with the Hamiltonian, and therefore, their eigenvalues can be used to label the system solutions. The solution can be found using the strategy of Ref.~\cite{Ambrus:2015lfr} generalized to the $N_f = 2$ flavor fermions in the constant Polyakov loop background with $N_c = 3$ colors. Omitting the technical details that will be presented elsewhere~\cite{in_preparation}, we outline briefly below the most essential features of this procedure.

Using the cumulative label $j$ to denote the eigenvalues, we impose
\begin{align}
 H U_j & = \tilde{\omega}^a_j U_j\,, \qquad 
 J^z U_j = m_j U_j\,, \nonumber\\ 
 P^z U_j & = k_j U_j\,, \qquad\ \
 h U_j = \lambda_j U_j\,.
 \label{eq_cyl_modes_eigen}
\end{align}
Here the frequency $\tilde{\omega}^a_j$ is related to the Minkowski energy $E_j > 0$ in the laboratory reference frame as follows:
\begin{equation}
 \tilde{\omega}^a_j = \varsigma_j E_j - {\mathcal A}_{0;j} - \mu - \Omega m_j\,.
 \label{eq_rot_omega}
\end{equation}
The right-hand side corresponds to a solution of the Dirac equation~\eqref{eq_rot_Dirac_eq} for which 
the symbol $\varsigma_j = \pm 1$ distinguishes between particle (positive energy) and anti-particle (negative energy) solutions. The superscript $a$ indicates that the gauge field contribution corresponding to the Polyakov-loop background has been taken into account, $\tilde{\omega}^a_j = \tilde{\omega}_j - {\mathcal A}_{0;j}$. The tilde notation denotes the subtraction of the angular momentum term, $\tilde{\omega}_j = \omega_j - \Omega m_j$, corresponding to the shift in the Hamiltonian~\eqref{eq_H_Hlab}, while $\omega_j = \varsigma_j E_j - \mu$. Notice that the index $m_j$ runs over half-integer numbers, $m_j = \pm 1/2, \pm 3/2, \dots$.

In the mean-field approximation that we are employing, the gauge field is diagonal in color space, $\mathcal{A}_0 = \frac{1}{2} (A^3_0 t^3 + A^8_0 t^8)$. The cylindrical modes~$U_j$ in Eq.~\eqref{eq_cyl_modes_eigen} can be taken as eigenvectors of both diagonal generators $t^3$ and $t^8$ of the gauge $SU(3)$ group. Thus, 
\begin{equation}
 \mathcal{A}_0 U_j = {\mathcal A}_{0;j} U_j,
 \label{eq_cyl_A0eigen}
\end{equation}
where we denote the components of the background field $(\mathcal{A}_{0;1}, {\mathcal A}_{0;2}, {\mathcal A}_{0;3}) = (\phi, \phi', -\phi - \phi')$,
\begin{align}
 \phi & = \frac{1}{2} \mathcal{A}^3_0 + \frac{1}{2\sqrt{3}} \mathcal{A}^8_0\,, \nonumber \\  
 \phi' & = -\frac{1}{2} \mathcal{A}^3_0 + \frac{1}{2\sqrt{3}} \mathcal{A}^8_0\,, \nonumber \\ 
 -\phi-\phi' & = -\frac{1}{\sqrt{3}} \mathcal{A}^8_0\,,
 \label{eq_cyl_A0eigen_aux}
\end{align}
of the SU(3) Polyakov loop:
\begin{align}
    L_+ & \equiv L = \frac{1}{3} \bigl(e^{\beta\phi} + e^{\beta \phi'} + e^{-\beta(\phi + \phi')}\bigr)\,, \\
    L_- & \equiv L^* = \frac{1}{3} \bigl(e^{-\beta\phi} + e^{-\beta \phi'} + e^{\beta(\phi + \phi')}\bigr)\,. 
\end{align}

The modes $U_j$ must satisfy boundary conditions on the enclosing cylinder of radius $R$. In this paper, we employ the spectral boundary conditions~\cite{Hortacsu:1980kv} as discussed for the cylindrical setup in Ref.~\cite{Ambrus:2015lfr}. These conditions provide necessary and sufficient conditions to yield a consistent quantization by imposing the self-adjointness of the Hamiltonian. In terms of the eigenspinors $U_j$, the spectral condition amounts to the vanishing of either the
top and third (for $m_j >0$) or the second and fourth components of the spinor $U_j$ (for $m_j <0$). These conditions guarantee the conservation of the total vector (baryon) charge inside the cylinder, as well as the self-adjointness of the Hamiltonian.

The solutions of the eigenvalue equations~\eqref{eq_cyl_modes_eigen} for a single-species fermion satisfying the spectral boundary conditions were reported in Ref.~\cite{Ambrus:2015lfr}. For our problem, the eigenenergies are given by Eq.~\eqref{eq_rot_omega}, where $E_j = \sqrt{p_j^2 + M^2}$ is the Minkowski energy, $M = g\sigma$ is the particle mass, $p_j^2 = \sqrt{q_j^2 + k_j^2}$ is the total momentum magnitude and the transverse momentum $q_j$ is quantized due to the imposed cylindrical boundary conditions:
\begin{equation}
 q_j R = \begin{cases}
  \xi_{m_j - \frac{1}{2}, \ell_j}, & m_j > 0\,, \\
  \xi_{-m_j - \frac{1}{2}, \ell_j}, & m_j < 0\,.
 \end{cases}
 \label{eq_cyl_spectral}
\end{equation}
Here $\xi_{n\ell}$ is the $\ell$th nonzero root of the Bessel function:
\begin{align}
    J_n(\xi_{n\ell}) = 0\,, \qquad {\rm with} \qquad n \ge 0\,, 
\end{align}
where the natural number $l = 1,2,\dots$ labels the radial modes. 

The quantization is compatible with the charge conjugation operation, by which the spinors $V_j(x) = i \gamma^2 U^*_j(x) = i(-1)^{m_j} \varsigma_j U_{\bar{\jmath}}(x)$ must also belong to the set of the modes $\{U_j\}$. Here, the cumulative index $\bar{\jmath}$ represents the set of charge-conjugated eigenvalues with respect to $j$, i.e.
\begin{align}
 j & = (\varsigma_j, f_j, c_j, \lambda_j, m_j, \ell_j, k_j), \nonumber \\
 \bar{\jmath} & = (-\varsigma_j, f_j, c_j, \lambda_j, -m_j, \ell_j, -k_j),
\end{align}
with $f_j = u,d$ and $c_j = 1,2,3$ being the flavour and colour quantum numbers, respectively.

\subsection{Free energy of the rotating system}

The average thermodynamic potential of the rotating fermionic ensemble can be split into a vacuum and a thermal part, similar to the static case considered above. The thermal part can be written in analogy with Eq.~\eqref{eq_P}:
\begin{align}
 F_{\psi\bar{\psi}}  = -\frac{2 T N_f}{\pi R^2}
 \sum_{\mathrm{b},\varsigma=\pm 1}\int_{-\infty}^\infty \frac{dk}{2\pi} \Tilde{F}_\varsigma\,,
 \label{eq_cyl_F}
\end{align}
where we introduced for notational brevity the sum over the radial quantum numbers,
\begin{equation}
    \sum_{\mathrm{b}} = \sum_{m = -\infty}^\infty \sum_{l = 1}^\infty\,,
    \label{eq_radial}
\end{equation}
keeping in mind that the index $m = \pm \frac{1}{2}, \pm \frac{3}{2}, \dots$ runs over all odd half-integer values.

In order to compute $\tilde{F}_\varsigma$ in Eq.~\eqref{eq_cyl_F}, we note the following relation: $\varsigma_j \tilde{\omega}^a_j = E_j - \varsigma_j (\Omega m_j + \mu + {\mathcal A}_{0;c})$. Flipping the sign of $m_j$ in the sum over ``$\mathrm{b}$'' for the antiparticle sector (when $\varsigma_j = -1$), we have $\varsigma_j \tilde{\omega}^a_j \rightarrow \widetilde{\mathcal{E}}_j - \varsigma_j {\mathcal A}_{0;c_j}$, with $\widetilde{\mathcal{E}}_j = \mathcal{E}_j - \Omega m_j$ being the corotating effective energy and $\mathcal{E}_j = E_j - \varsigma_j \mu$. Then, $\widetilde{F}_\varsigma$ can be computed using the colour eigenvalue Eqs.~\eqref{eq_cyl_A0eigen}--\eqref{eq_cyl_A0eigen_aux} as follows:
\begin{align}
 \widetilde{F}_\varsigma & = \sum_{c = 1}^3 \ln(1 + e^{-\beta (\widetilde{\mathcal{E}}_\varsigma - \varsigma {\mathcal A}_{0;c})}) \nonumber\\
 & = \ln[1 + 3 L_\varsigma e^{-\beta \widetilde{\mathcal{E}}_\varsigma} + 3 L_{-\varsigma} e^{-2\beta \widetilde{\mathcal{E}}_\varsigma} + e^{-3\beta \widetilde{\mathcal{E}}_\varsigma}]\,.
 \label{eq_cyl_Fs}
\end{align}
Notice that $\widetilde{F}_\varsigma$ in Eq.~\eqref{eq_cyl_Fs} coincides with the one in Eq.~\eqref{eq_F_psi} after replacing the laboratory-frame energy $\mathcal{E}_\varsigma$ with the corotating energy, $\widetilde{\mathcal{E}}_\varsigma$. 

The thermal free energy  $F_{\psi\bar{\psi}}$ in Eq.~\eqref{eq_cyl_F} is identical to the one for the static reference frame~\eqref{eq_FQuark} after replacing the integration with respect to the three-momentum via 
\begin{equation}
 \int \frac{d^3p}{(2\pi)^3} \rightarrow \frac{1}{\pi R^2} \sum_{\mathrm{b}} \int \frac{dk}{2\pi}\,.
 \label{eq_rot_int_measure}
\end{equation}

So far, in this section, we have only discussed the thermodynamics of the fermionic part of the model. The bosonic part is encoded in the classical condensate $\sigma$, which is not affected by rotation directly. The effect of rotation on the field $\sigma$ appears only through the interaction with fermionic loops, which do feel the rotation directly, as we discussed in this section. Moreover, we treat the condensate $\sigma$ in the homogeneous approximation that is used in most of the studies so far. Namely, the condensate is set to be a coordinate-independent quantity. While noticing that the homogeneous approximation allows us to compare our results with the results of other approaches, we acknowledge that the realistic rotating plasma should develop a radial inhomogeneity, which should become rather pronounced for fast rotation~\cite{Chernodub:2020qah, Chernodub:2022veq, Braguta:2023iyx}. 
We will address the inhomogeneous thermodynamic ground state of the  vortical plasma elsewhere.

\section{Splitting of transitions}
\label{sec_split}

In this section, we thoroughly evaluate the phase diagram of the PLSM${}_q$ inside the cylindrical volume following our strategy described in Section~\ref{sec_rotating}. 
While the static PLSM has been intensively investigated in the literature, the effects of the cylindrical boundaries on the phase diagram have never been studied in this model. Noticing that the boundary conditions are essential for our study to keep causality in place, we investigate the effect of boundaries on the splitting of the chiral and deconfinement transitions below.

\subsection{Effect of cylindrical boundaries for static plasma}

\subsubsection{Softening, shifting, and splitting of transitions}

We minimize the free energy of the system enclosed in the cylindrical cavity, Eqs.~\eqref{eq_F} with the fermionic free energy given in Eq.~\eqref{eq_cyl_F}. The latter quantity can only be evaluated with the use of extensive numerical methods since the sum over the radial excitations~\eqref{eq_radial}, provided by the zeros of Bessel functions~\eqref{eq_cyl_spectral}, can involve up to $10^5$ terms to guarantee the convergence of the sum.

\begin{figure*}[!htb]
\centering
\begin{tabular}{cc}
\includegraphics[width=0.49\linewidth]{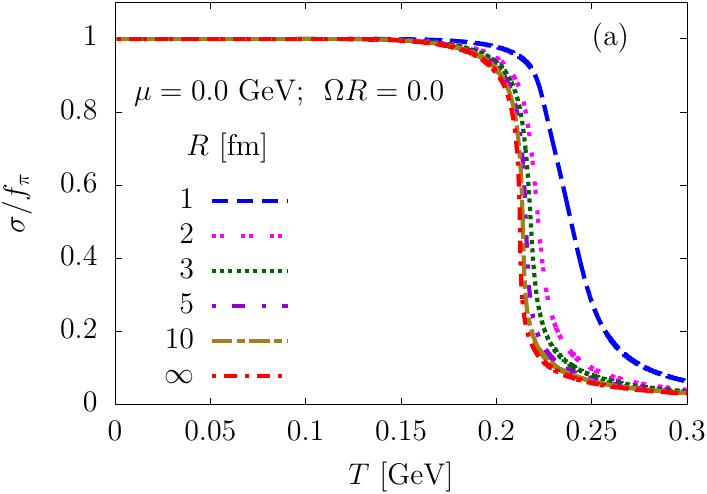} &
\includegraphics[width=0.49\linewidth]{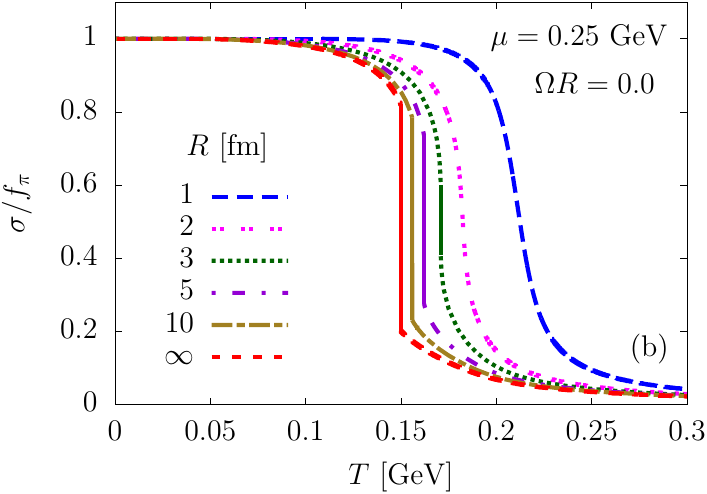} \\
\includegraphics[width=0.49\linewidth]{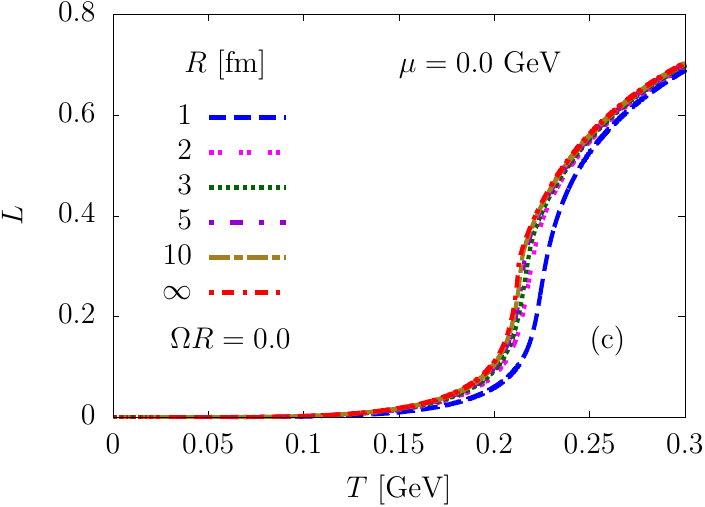} &
\includegraphics[width=0.49\linewidth]{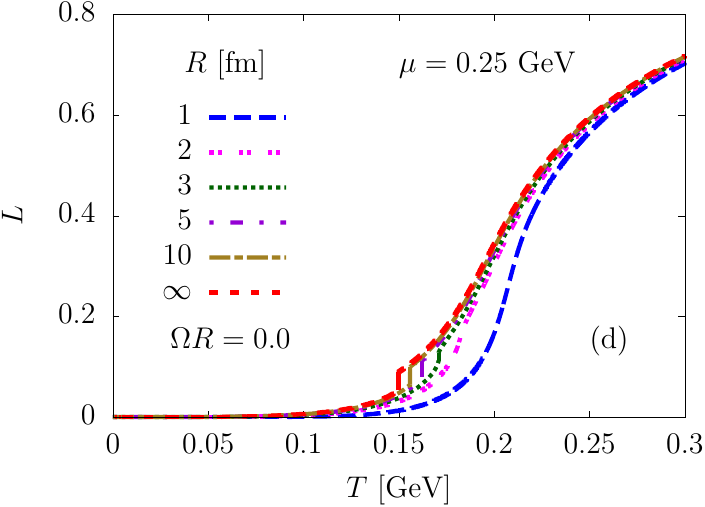}
\end{tabular}
\caption{The dependence on temperature of (top) the normalized chiral order parameter $\sigma/f_\pi$ and (bottom) the confinement order parameters $L$ and $L^*$ for a static ($\Omega R = 0$) cylindrical system of various radii for (left) vanishing, $\mu = 0$ and (right) a non-zero, $\mu = 0.25$~GeV, chemical potentials.
}
\label{fig_rot0_sL}
\end{figure*}

First, we consider the non-rotating case with $\Omega = 0$. In Fig.~\ref{fig_rot0_sL}, we show the mean values of the chiral condensate and the Polyakov loops as functions of temperature for various radii $R$ of the cylinder at chemical potentials $\mu = 0$ and $\mu = 0.25$~GeV. In an infinite volume and for vanishing chemical potential, $\mu=0$, the transition has the nature of a smooth crossover, since none of the order parameters show anything similar to critical behavior. At $\mu = 0.25$~GeV, the $R \to \infty$ transition is a strong first-order phase transition. What happens with these transitions when we decrease the radius of the cylinder~$R$?

First, we notice that the decrease of the radius leads to the softening of the transition seen in the behaviour of all order parameters and for all chemical potentials. This property agrees with our experience for a finite volume, for which -- if the system is bounded in all three directions -- the thermodynamic phase transition cannot be realized. Therefore, even a first-order phase transition (that takes place at higher chemical potential) is softened, becoming a crossover transition in a finite volume. An existing smooth crossover at zero chemical potential becomes even softer. A similar softening behavior of our system, for which only two dimensions out of three are restricted, is compatible with the mentioned finite-size softening effect.

The described softening of the transition temperature is most clearly seen for the condensate $\sigma$ 
for the large chemical potential $\mu = 0.25$~GeV. As the radius diminishes, the jump in the condensate diminishes and, at $R = 2$~fm, it is not seen at all. This radius corresponds to the energy scale of 100~MeV, which is compatible with the pion mass. Therefore, the restriction of one of the dimensions to this value leads to a drastic softening effect on the phase transition. 

Second, we find that the decrease in the radius of the cylinder $R$ leads to an increase in the pseudocritical temperatures of both crossovers. This effect can also be anticipated given the fact that the decrease of the radius leads to a suppression of the thermal contribution to the chiral condensate inside the cylinder~\cite{Ambrus:2015lfr}, due to an enhancement of the mass gap induced by the transverse momentum quantization. In order to overcome this gap, higher temperatures are needed. Due to the coupling between the chiral and confining degrees of freedom in the PLSM${}_q$, the expectation value of the Polyakov loop follows the same behavior. The enhancement of the critical temperature is, therefore, a finite-volume phenomenon.

Third, we see that the inflection points at the smallest studied radius $R = 1$~fm of the $\sigma$ condensate and the mean Polyakov loop do not coincide, with the deconfinement crossover transition being slightly lower than the chiral transition. Therefore, one expects that in a system with bounded spatial dimensions, the chiral and confinement transitions split, thus creating an already-deconfined but still-chirally broken quark-gluon matter. 

Strictly speaking, due to the crossover nature of these transitions and a finite width of both transitions, the confinement and chiral restoration never set exactly at the mentioned intermediate region of temperatures. Nevertheless, this intermediate phase is characterized by a nonzero $\sigma$ condensate (a chirally broken regime) and a nonzero expectation value of the Polyakov loop $L$ (a deconfinement regime). This splitting effect produced by the finite size of the system becomes stronger for a denser system.

Thus, the softening, shifting and splitting of the chiral and deconfining transitions are finite-volume effects.

\subsubsection{Quantifying the splitting of transition temperatures}

\begin{figure}[!htb]
\centering
\begin{tabular}{c}
\includegraphics[width=0.99\linewidth]{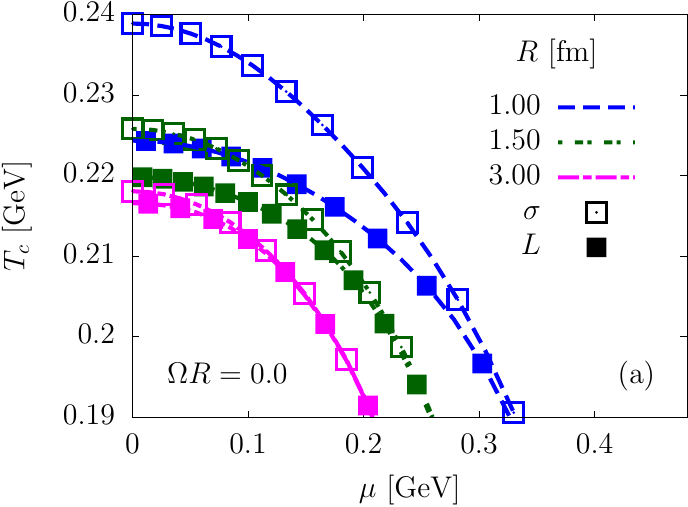} \\
\includegraphics[width=0.99\linewidth]{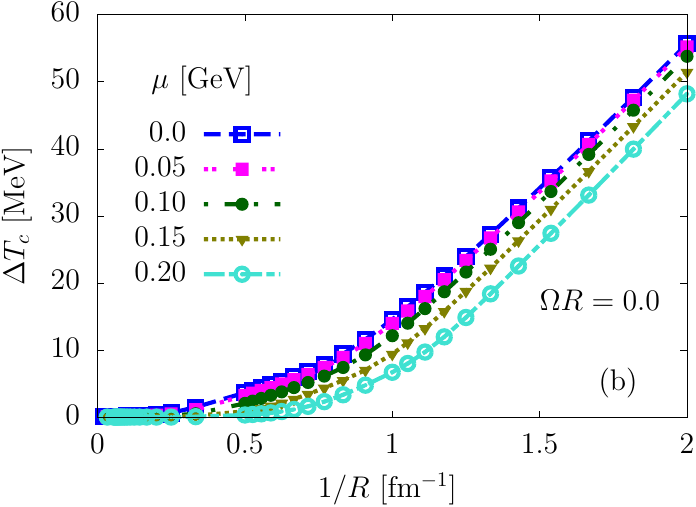} 
\end{tabular}
\caption{
(a) The crossover temperatures as a function of the chemical potential $\mu$ for the chiral transition and the deconfinement transition determined from the inflection points of the $\sigma$ condensate and $L$ (or $L^*$). Several values of the radius of the cylinder are shown. (b) The split in the transition temperatures~\eqref{eq_Delta_T_def} for chiral and confinement-deconfinement crossovers at different fixed chemical potentials as a function of the (inverse) radius. All plots are shown for zero angular velocity, $\Omega R = 0$.}
\label{fig_rot0_PD}
\end{figure}

How strong is the splitting in chiral and deconfining temperatures? To this end, we found the inflection points in the chiral order parameter, $\sigma$, and in the deconfining order parameter, $L$. We plot these crossover temperatures in Fig.~\ref{fig_rot0_PD}(a) at a fixed radius of the system $R$ determined along a diagonal line crossing the origin of the parameter space, $T = \mu = 0$.

Figure~\ref{fig_rot0_PD}(a) demonstrates that the splitting takes its maximal value at vanishing chemical potential, $\mu =0$. As the chemical potential increases, the splitting decreases. Thus, the finite-volume effect of the splitting gets reduced in the finite-density system. Moreover, we see the already noticed property that the decrease in the radius of the cylinder increases the splitting.

The finite-volume splitting effect is rather noticeable at $R = 1$~fm, where the difference in temperature achieves a rather modest $5-6$\% of the pseudo-critical temperature. However, in this case, the radius of the cylinder is about the size of the hadron, so that at this size, strictly speaking, the system is not in a bulk state, at least with respect to the transverse directions. In other words, there is space only for two or three hadrons to fit into the transverse plane of the area $A_\perp \simeq 3.14 \,{\rm fm}^2$, which does not allow us to consider this system as a thermodynamically large ensemble, at least, in the transverse directions. 

However, already at $R = 3$~fm, the splitting of the chiral and deconfining transitions is barely noticeable. The transverse area becomes rather large, $A_\perp \simeq 28 \,{\rm fm}^2$, which allows us to consider this system as a bulk medium that gets closer to the thermodynamic limit, $R \to \infty$. And, in agreement with almost no splitting found at $R = 3$~fm, there is no splitting between the transitions in the thermodynamic limit, as we have found in the previous section.

Figure~\ref{fig_rot0_PD}(b) allows us to quantify the magnitude of the splitting as a function of the radius. Here we evaluate the crossover temperatures along the line of constant $\mu$ and for presentation reasons, we plot the split in the pseudo-critical temperatures of chiral, $T_{c}^{(\sigma)}$, and deconfining, $ T_{c}^{(L)}$, transitions:
\begin{align}
    \Delta T_{c} = T_{c}^{(\sigma)} - T_{c}^{(L)} > 0\,,
    \label{eq_Delta_T_def}
\end{align}
as a function of the inverse radius, $1/R$. This Figure nicely illustrates two properties of the non-rotating system: both the increase in the radius of the cylinder and the increase of the baryon density of the medium inside the cylinder diminish the split. For the discussed $R = 3$~fm cylinder, which can loosely be considered a bulk system, the finite-size splitting amounts to a tiny $\Delta T_{pc} \sim 1$~MeV at zero-density plasma ($\mu = 0$). Notice that to maintain consistency in our notation, we denote the pseudo-critical temperatures $T_{pc}$ also by $T_c$.

After determining the effect of the finite volume on the splitting of the transition temperatures, we can look at the impact of rotation on the bulk properties of the system, such as the phase transition diagram. Here, the word ``bulk'' implies that we neglect the surface effect of the boundary related to the existence of the mass gap of the system. In a system with a mass gap $M$, the boundary affects the properties of the system given by the distance of one correlation length $\lambda = 1/M$ from the boundary. The phenomena found in this section so far can be considered as the manifestation of this edge effect.

\subsection{Effect of rotation on temperature splitting}

\begin{figure}[!htb]
\centering
\begin{tabular}{c}
\includegraphics[width=0.98\linewidth]{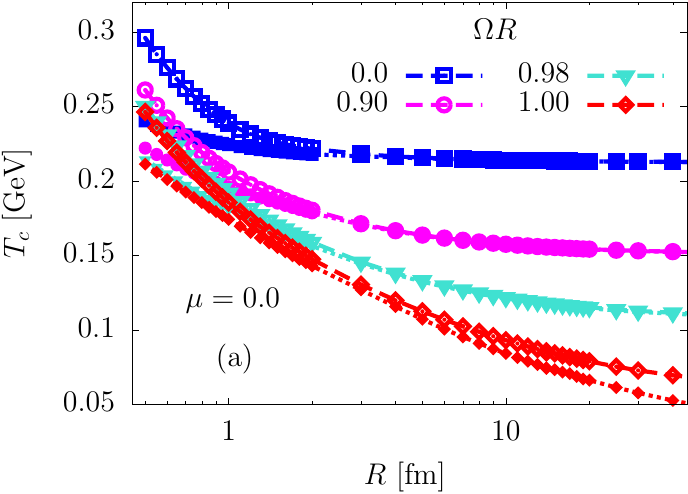} \\
(a) \\
\includegraphics[width=0.98\linewidth]{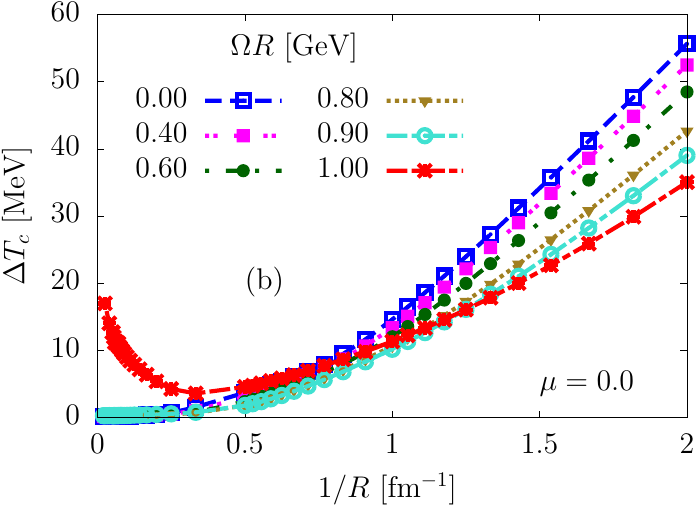} \\
(b)
\end{tabular}
\caption{
Effect of rotation on the separation of chiral and deconfinement crossovers: (a) the pseudocritical temperatures as a function of $R$ and (b) their difference $\Delta T_c$ \eqref{eq_Delta_T_def} as functions of the cylinder radius $R$ and its inverse, $1/R$, respectively, at different values of $\Omega R$. The results are shown for vanishing chemical potential $\mu = 0$, which gives the maximum temperature difference $\Delta T_c$. In panel (a), the empty (filled) symbols correspond to chiral (deconfinement) transitions.
}
\label{fig_Delta_T_mu0}
\end{figure}

We now consider the pseudocritical transition temperatures $T_{pc}^{(\sigma)}$ and $T_{pc}^{(L)}$ at finite volume and with rotation for the case when $\mu = 0$, when the split \eqref{eq_Delta_T_def} between these two temperatures is maximal, as shown in Fig.~\ref{fig_Delta_T_mu0}. Panel (a) of Fig.~\ref{fig_Delta_T_mu0} shows the dependence of both $T_c(L)$ (upper, dashed lines with empty symbols) and $T_c(\sigma)$ (the lower, solid lines with the filled symbols) on the radius of the cylindrical boundary, for various values of the rotation parameter, taken such that $\Omega R = 0$ (the blue lines and the rhombi), $0.9$ (the orange lines and the triangles), $0.98$ (the green lines and the circles) and $1$ (purple lines and squares) is kept fixed. When $\Omega R < 1$, it can be seen that the gap $T_{pc}^{(L)} - T_{pc}^{(\sigma)}$ reduces as $R$ is increased, becoming negligible when $R \gtrsim 2$ fm. In the case when $\Omega R = 1$, the gap never fully disappears.

In panel (b) of Figure~\ref{fig_Delta_T_mu0}, we show the difference $\Delta T_{pc}$ of the transition temperatures~\eqref{eq_Delta_T_def} as a function of the inverse radius of the cylinder, $1/R$, at various rotation frequencies $\Omega R$. The results are shown for vanishing chemical potential, $\mu = 0$, where the splitting takes its maximum value. One can clearly see that the rotation does not increase the splitting. On the contrary, the difference in temperatures between the chiral and deconfining transitions, generated by the finite-volume effects, becomes even smaller as the rotation frequency increases. 

Thus, rotation inhibits the splitting of temperatures rather than producing it.

Let us relate the results shown in Fig.~\ref{fig_Delta_T_mu0} to the physical conditions corresponding to the realistic conditions of the vortical plasmas produced in the relativistic heavy-ion collisions. In the RHIC experiment, the vorticity of the quark-gluon plasma corresponds to the angular frequency of $\Omega \simeq 6.6$~MeV~\cite{STAR:2017ckg}, which translates to $\Omega R \simeq 0.165$ for a system of a rather large size $R = 5$~fm. Thus, the rotation velocity at the boundary of the system is much smaller than the ultrarelativistic limit $\Omega R = 1$. To put this number in a different perspective, the rotational Lorentz factor at the edge of the system for RHIC-type collisions amounts to $1/\sqrt{1 - (\Omega R)^2} \approx 1.01$, providing us with a minuscule one-percent enhancement. In other words, the effect of the presence of the boundary of a rotating system on the thermodynamic properties of this system is much larger than the effect of the rotation itself.

\begin{figure*}[!htb]
\centering
\begin{tabular}{cc}
\includegraphics[width=0.49\linewidth]{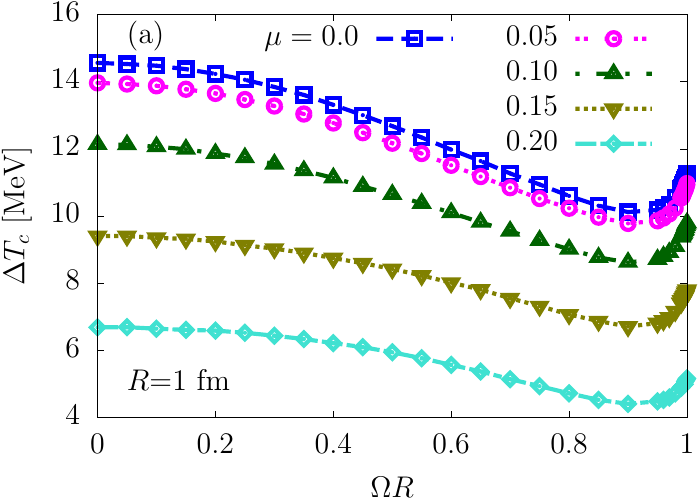}&
\includegraphics[width=0.49\linewidth]{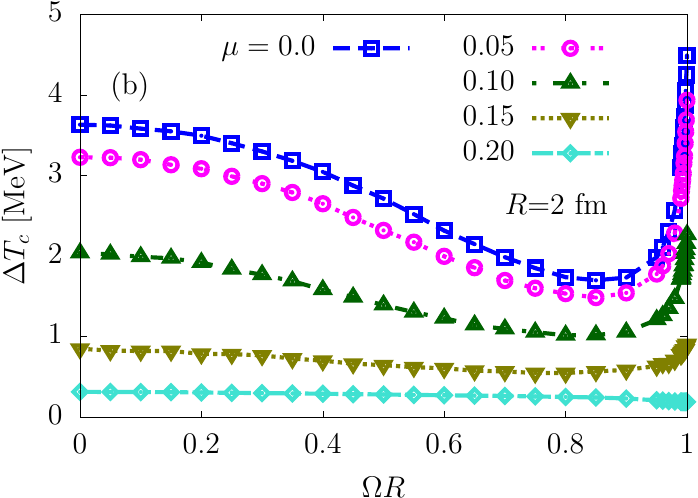} 
\\
\includegraphics[width=0.49\linewidth]{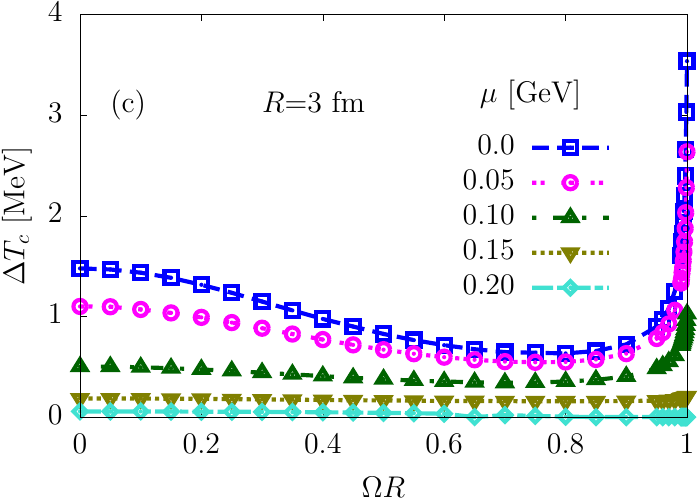}&
\includegraphics[width=0.49\linewidth]{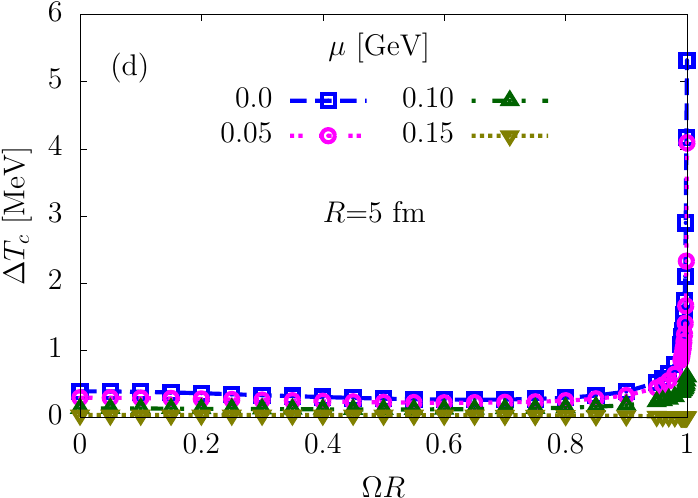}
\end{tabular}
\caption{Difference in the crossover temperatures $\Delta T_c$ for chiral and confinement-deconfinement transitions at different values of the chemical potential $\mu$ as a function of angular velocity, given in units of $\Omega R$. Each plot corresponds to a fixed radius of the cylinder, $R = (1, 2, 3, 5)$~fm}
\label{fig_Delta_T}
\end{figure*} 

Consider also an academic limit of the ultrarelativistic vorticities, for which the parameter $\Omega R$ reaches $\Omega R = 1$ value. Since at such rotation, the matter effects may potentially be also important, in Fig.~\ref{fig_Delta_T}, we plot the splitting as the function of $\Omega R$ for various values of the chemical potential $\mu$ at different sizes $R$ of the system. The results confirm that, similarly to the finite-volume effects, the increasing chemical potential inhibits the split in the transition temperatures. Moreover, for all chemical potential, the increase of the rotation frequency up to $\Omega R \simeq 0.9$ inhibits the temperature difference. However, at event faster rotations, as $\Omega R \to 1$, the splitting in the transition temperature starts to increase, reaching a larger value at $\Omega R = 1$. Still, even at $\mu = 0$, the rotation-induced splitting in the academic ultrarelativistic regime reaches a rather moderate value, $\Delta T_c \simeq 11$~MeV.


\begin{figure}
\begin{tabular}{c}
\includegraphics[width=.98\linewidth]{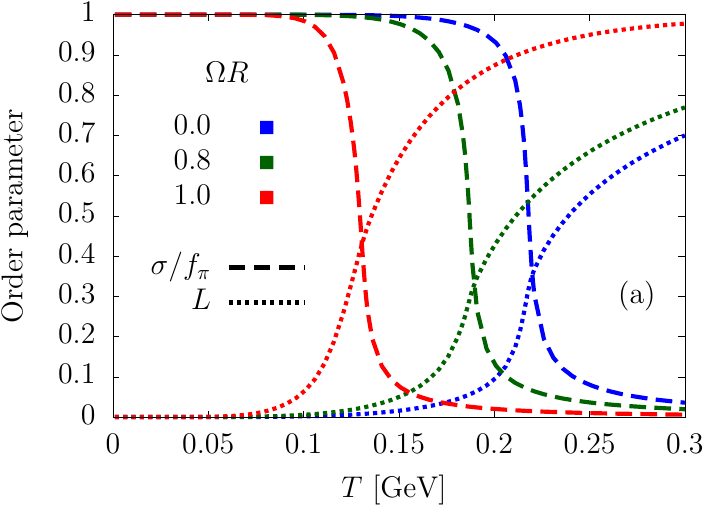} \\
\includegraphics[width=.98\linewidth]{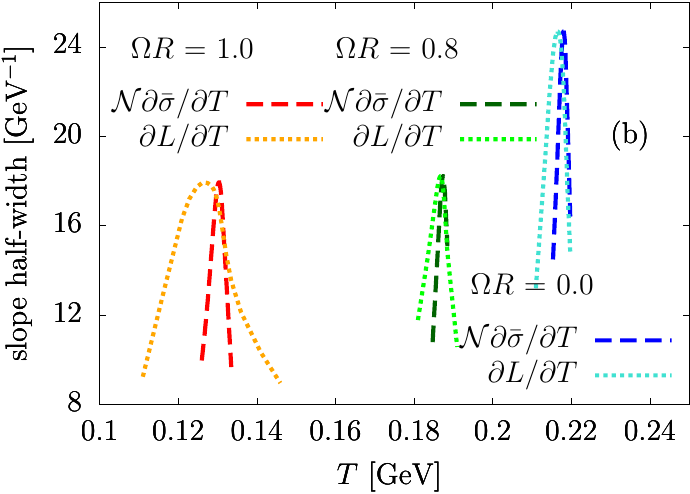}
\end{tabular}
\caption{Variation of (a) the $\sigma$ and $L$ order parameters; and (b) the slopes $\partial \bar{\sigma} / \partial T$ $(\bar{\sigma}\equiv \sigma/f_\pi)$ and $\partial L / \partial T$, with respect to the temperature $T$, for a system of size $R = 3$ fm, at various values of $\Omega R$ at $\mu=0$. The slope $d\bar{\sigma}/dT$ is scaled by a constant factor $\mathcal{N}$ for better visibility.}
\label{fig:overlap}
\end{figure}

According to Fig.~\ref{fig_Delta_T}, the split in the transition temperature appears to grow when $\Omega R$ approaches the maximal value limited by the causality, $\Omega R = 1$. 
As can be seen in panel (a) of Fig.~\ref{fig_Delta_T_mu0}, the gap $\Delta T_c (\Omega R = 1)$  initially decreases with increasing $R$, while when $R$ increases beyond $3$~fm, it increases. Since the split appears due to a lower pseudocritical temperature for the Polyakov loop $L$ than for $\sigma$, we can expect that the gap $\Delta T_c$ widens with increasing $R$ and then again shrinks, as for both $L$ and $\sigma$, $T_c \rightarrow 0$ when $R \rightarrow \infty$. We note, however, that probing this large-$R$ regime is computationally expensive, and we therefore leave it for the more detailed analysis in our forthcoming publication \cite{in_preparation}.

Finally, we would like to comment on the nature of the crossover transition in the regime where the chiral and deconfinement transition temperatures do not coincide. Panel (a) of Fig.~\ref{fig:overlap} shows the temperature dependence of both the sigma meson $\sigma$ and the Polyakov loop $L$, in the case when $R = 3$ fm and $\mu = 0$, for $\Omega R = 0$, $0.8$ and $1$. These values of $\Omega R$ were chosen to correspond to the maximum value of $\Delta T_{pc}$, its minimum value, and again the peak reached at the causality limit, respectively. Looking at the plot, we can see that the chiral and deconfinement transitions are very much overlapped and no clear distinction between their respective pseudocritical transition temperatures can be seen. Panel (b) of the same figure shows the scaled slopes $\mathcal{N}\partial \bar{\sigma} / \partial T$ and $\partial L / \partial T$, where $\bar{\sigma}\equiv \sigma/f_\pi$, for the above mentioned values of $\Omega R$. Here the normalization $\mathcal{N}$ is such that the peak of $\partial \bar{\sigma} / \partial T$ lies at the same value as that of $\partial L / \partial T$. While the peaks of these slopes, used by us to identify the corresponding pseudocritical temperature, do not coincide, the bell-shaped areas developing around these peaks are overlapped. We can thus conclude that a discussion of a unique value for the pseudocritical temperature, and hence the identification of a split between deconfinement and chiral symmetry restoration, is at best ambiguous in the case of bounded rotating systems.

\section{Conclusions}

In our paper, we discussed how the rigid rotation of quark-gluon plasma, modeled by the Polyakov-loop enhanced linear sigma model with quarks, affects the critical temperatures of deconfinement and chiral transitions. We highlighted the importance of the causality condition, which requires that any point in the system rotates slower than the speed of light. Since the rotation is rigid, the causality condition limits the size of the system in the transverse directions, $R < 1/\Omega$, where $\Omega>0$ is the angular frequency.

We found that the confinement and chiral transitions split in the system. However, this splitting effect originates not due to the rotation of the system but rather due to its finite volume. In a very small non-rotating system with a radius of one fermi, the splitting can reach a modest $\Delta T_c \sim 10$~MeV, which is comparable with the width of the crossover transition~\cite{Aoki:2006br, Borsanyi:2010bp, Bazavov:2011nk}. For systems of the size of about 3-5~fm, the splitting becomes very negligible, $\Delta T \sim 1$~MeV. 

Contrary to expectations, the rotation appears to inhibit this finite-volume splitting: as the angular velocity increases, the chiral and deconfining temperatures approach each other. The splitting is also inhibited by a finite chemical potential: as the density of the plasma increases, the splitting between the chiral and confining transitions becomes smaller. 

We have also considered an ultra-relativistic regime where the boundary of the system rotates very close to the speed of light. In this, the temperature splitting becomes visible again, but it still remains in the modest 10~MeV range. 

One should stress that we used the uniform approximation, in which both the chiral condensate and the expectation value of the Polyakov loop are treated as coordinate-independent quantities. This approach, which can be justified for the central region of the plasma that rotates with a slow rotational velocity, $\Omega R \ll 1$, cannot be applied for rapidly rotating systems for which the boundary velocity reaches $\Omega R \sim 1$. For such plasmas, the inhomogeneities must be taken into account. 
This important question will be addressed elsewhere. Nevertheless, we believe that the results of this paper, obtained in the uniform approximation, are applicable to realistic plasmas that rotate relatively slowly.

We conclude that rotation is unlikely to induce the splitting between chiral and confining transitions in realistic vortical plasmas created in relativistic heavy-ion collisions. 

\acknowledgments
This work is supported by the European Union - NextGenerationEU through grant No. 760079/23.05.2023, funded by the Romanian ministry of research, innovation and digitalization through Romania’s National Recovery and Resilience Plan, call no. PNRR-III-C9-2022-I8. 

\bibliography{ref}

\begin{thebibliography}{63}%
\makeatletter
\providecommand \@ifxundefined [1]{%
 \@ifx{#1\undefined}
}%
\providecommand \@ifnum [1]{%
 \ifnum #1\expandafter \@firstoftwo
 \else \expandafter \@secondoftwo
 \fi
}%
\providecommand \@ifx [1]{%
 \ifx #1\expandafter \@firstoftwo
 \else \expandafter \@secondoftwo
 \fi
}%
\providecommand \natexlab [1]{#1}%
\providecommand \enquote  [1]{``#1''}%
\providecommand \bibnamefont  [1]{#1}%
\providecommand \bibfnamefont [1]{#1}%
\providecommand \citenamefont [1]{#1}%
\providecommand \href@noop [0]{\@secondoftwo}%
\providecommand \href [0]{\begingroup \@sanitize@url \@href}%
\providecommand \@href[1]{\@@startlink{#1}\@@href}%
\providecommand \@@href[1]{\endgroup#1\@@endlink}%
\providecommand \@sanitize@url [0]{\catcode `\\12\catcode `\$12\catcode
  `\&12\catcode `\#12\catcode `\^12\catcode `\_12\catcode `\%12\relax}%
\providecommand \@@startlink[1]{}%
\providecommand \@@endlink[0]{}%
\providecommand \url  [0]{\begingroup\@sanitize@url \@url }%
\providecommand \@url [1]{\endgroup\@href {#1}{\urlprefix }}%
\providecommand \urlprefix  [0]{URL }%
\providecommand \Eprint [0]{\href }%
\providecommand \doibase [0]{http://dx.doi.org/}%
\providecommand \selectlanguage [0]{\@gobble}%
\providecommand \bibinfo  [0]{\@secondoftwo}%
\providecommand \bibfield  [0]{\@secondoftwo}%
\providecommand \translation [1]{[#1]}%
\providecommand \BibitemOpen [0]{}%
\providecommand \bibitemStop [0]{}%
\providecommand \bibitemNoStop [0]{.\EOS\space}%
\providecommand \EOS [0]{\spacefactor3000\relax}%
\providecommand \BibitemShut  [1]{\csname bibitem#1\endcsname}%
\let\auto@bib@innerbib\@empty
\bibitem [{\citenamefont {Abelev}\ \emph {et~al.}(2007)\citenamefont {Abelev}
  \emph {et~al.}}]{STAR:2007ccu}%
  \BibitemOpen
  \bibfield  {author} {\bibinfo {author} {\bibfnamefont {B.~I.}\ \bibnamefont
  {Abelev}} \emph {et~al.} (\bibinfo {collaboration} {STAR}),\ }\bibfield
  {title} {\enquote {\bibinfo {title} {{Global polarization measurement in
  Au+Au collisions}},}\ }\href {\doibase 10.1103/PhysRevC.76.024915} {\bibfield
   {journal} {\bibinfo  {journal} {Phys. Rev. C}\ }\textbf {\bibinfo {volume}
  {76}},\ \bibinfo {pages} {024915} (\bibinfo {year} {2007})},\ \bibinfo {note}
  {[Erratum: Phys.Rev.C 95, 039906 (2017)]},\ \Eprint
  {http://arxiv.org/abs/0705.1691} {arXiv:0705.1691 [nucl-ex]} \BibitemShut
  {NoStop}%
\bibitem [{\citenamefont {Adamczyk}\ \emph {et~al.}(2017)\citenamefont
  {Adamczyk} \emph {et~al.}}]{STAR:2017ckg}%
  \BibitemOpen
  \bibfield  {author} {\bibinfo {author} {\bibfnamefont {L.}~\bibnamefont
  {Adamczyk}} \emph {et~al.} (\bibinfo {collaboration} {STAR}),\ }\bibfield
  {title} {\enquote {\bibinfo {title} {{Global $\Lambda$ hyperon polarization
  in nuclear collisions: evidence for the most vortical fluid}},}\ }\href
  {\doibase 10.1038/nature23004} {\bibfield  {journal} {\bibinfo  {journal}
  {Nature}\ }\textbf {\bibinfo {volume} {548}},\ \bibinfo {pages} {62--65}
  (\bibinfo {year} {2017})},\ \Eprint {http://arxiv.org/abs/1701.06657}
  {arXiv:1701.06657 [nucl-ex]} \BibitemShut {NoStop}%
\bibitem [{\citenamefont {Huang}\ \emph {et~al.}(2021)\citenamefont {Huang},
  \citenamefont {Liao}, \citenamefont {Wang},\ and\ \citenamefont
  {Xia}}]{Huang:2020dtn}%
  \BibitemOpen
  \bibfield  {author} {\bibinfo {author} {\bibfnamefont {Xu-Guang}\
  \bibnamefont {Huang}}, \bibinfo {author} {\bibfnamefont {Jinfeng}\
  \bibnamefont {Liao}}, \bibinfo {author} {\bibfnamefont {Qun}\ \bibnamefont
  {Wang}}, \ and\ \bibinfo {author} {\bibfnamefont {Xiao-Liang}\ \bibnamefont
  {Xia}},\ }\bibfield  {title} {\enquote {\bibinfo {title} {{Vorticity and Spin
  Polarization in Heavy Ion Collisions: Transport Models}},}\ }\href {\doibase
  10.1007/978-3-030-71427-7_9} {\bibfield  {journal} {\bibinfo  {journal}
  {Lect. Notes Phys.}\ }\textbf {\bibinfo {volume} {987}},\ \bibinfo {pages}
  {281--308} (\bibinfo {year} {2021})},\ \Eprint
  {http://arxiv.org/abs/2010.08937} {arXiv:2010.08937 [nucl-th]} \BibitemShut
  {NoStop}%
\bibitem [{\citenamefont {Becattini}\ \emph {et~al.}(2021)\citenamefont
  {Becattini}, \citenamefont {Liao},\ and\ \citenamefont
  {Lisa}}]{Becattini:2021lfq}%
  \BibitemOpen
  \bibfield  {author} {\bibinfo {author} {\bibfnamefont {Francesco}\
  \bibnamefont {Becattini}}, \bibinfo {author} {\bibfnamefont {Jinfeng}\
  \bibnamefont {Liao}}, \ and\ \bibinfo {author} {\bibfnamefont {Michael}\
  \bibnamefont {Lisa}},\ }\bibfield  {title} {\enquote {\bibinfo {title}
  {{Strongly Interacting Matter Under Rotation: An Introduction}},}\ }\href
  {\doibase 10.1007/978-3-030-71427-7_1} {\bibfield  {journal} {\bibinfo
  {journal} {Lect. Notes Phys.}\ }\textbf {\bibinfo {volume} {987}},\ \bibinfo
  {pages} {1--14} (\bibinfo {year} {2021})},\ \Eprint
  {http://arxiv.org/abs/2102.00933} {arXiv:2102.00933 [nucl-th]} \BibitemShut
  {NoStop}%
\bibitem [{\citenamefont {Yamamoto}\ and\ \citenamefont
  {Hirono}(2013)}]{Yamamoto:2013zwa}%
  \BibitemOpen
  \bibfield  {author} {\bibinfo {author} {\bibfnamefont {Arata}\ \bibnamefont
  {Yamamoto}}\ and\ \bibinfo {author} {\bibfnamefont {Yuji}\ \bibnamefont
  {Hirono}},\ }\bibfield  {title} {\enquote {\bibinfo {title} {{Lattice QCD in
  rotating frames}},}\ }\href {\doibase 10.1103/PhysRevLett.111.081601}
  {\bibfield  {journal} {\bibinfo  {journal} {Phys. Rev. Lett.}\ }\textbf
  {\bibinfo {volume} {111}},\ \bibinfo {pages} {081601} (\bibinfo {year}
  {2013})},\ \Eprint {http://arxiv.org/abs/1303.6292} {arXiv:1303.6292
  [hep-lat]} \BibitemShut {NoStop}%
\bibitem [{\citenamefont {Braguta}\ \emph {et~al.}(2020)\citenamefont
  {Braguta}, \citenamefont {Kotov}, \citenamefont {Kuznedelev},\ and\
  \citenamefont {Roenko}}]{Braguta:2020biu}%
  \BibitemOpen
  \bibfield  {author} {\bibinfo {author} {\bibfnamefont {V.~V.}\ \bibnamefont
  {Braguta}}, \bibinfo {author} {\bibfnamefont {A.~Yu.}\ \bibnamefont {Kotov}},
  \bibinfo {author} {\bibfnamefont {D.~D.}\ \bibnamefont {Kuznedelev}}, \ and\
  \bibinfo {author} {\bibfnamefont {A.~A.}\ \bibnamefont {Roenko}},\ }\bibfield
   {title} {\enquote {\bibinfo {title} {{Study of the Confinement/Deconfinement
  Phase Transition in Rotating Lattice SU(3) Gluodynamics}},}\ }\href {\doibase
  10.31857/S1234567820130029} {\bibfield  {journal} {\bibinfo  {journal} {Pisma
  Zh. Eksp. Teor. Fiz.}\ }\textbf {\bibinfo {volume} {112}},\ \bibinfo {pages}
  {9--16} (\bibinfo {year} {2020})}\BibitemShut {NoStop}%
\bibitem [{\citenamefont {Braguta}\ \emph {et~al.}(2021)\citenamefont
  {Braguta}, \citenamefont {Kotov}, \citenamefont {Kuznedelev},\ and\
  \citenamefont {Roenko}}]{Braguta:2021jgn}%
  \BibitemOpen
  \bibfield  {author} {\bibinfo {author} {\bibfnamefont {V.~V.}\ \bibnamefont
  {Braguta}}, \bibinfo {author} {\bibfnamefont {A.~Yu.}\ \bibnamefont {Kotov}},
  \bibinfo {author} {\bibfnamefont {D.~D.}\ \bibnamefont {Kuznedelev}}, \ and\
  \bibinfo {author} {\bibfnamefont {A.~A.}\ \bibnamefont {Roenko}},\ }\bibfield
   {title} {\enquote {\bibinfo {title} {{Influence of relativistic rotation on
  the confinement-deconfinement transition in gluodynamics}},}\ }\href
  {\doibase 10.1103/PhysRevD.103.094515} {\bibfield  {journal} {\bibinfo
  {journal} {Phys. Rev. D}\ }\textbf {\bibinfo {volume} {103}},\ \bibinfo
  {pages} {094515} (\bibinfo {year} {2021})},\ \Eprint
  {http://arxiv.org/abs/2102.05084} {arXiv:2102.05084 [hep-lat]} \BibitemShut
  {NoStop}%
\bibitem [{\citenamefont {Braguta}\ \emph
  {et~al.}(2023{\natexlab{a}})\citenamefont {Braguta}, \citenamefont {Kotov},
  \citenamefont {Roenko},\ and\ \citenamefont {Sychev}}]{Braguta:2022str}%
  \BibitemOpen
  \bibfield  {author} {\bibinfo {author} {\bibfnamefont {V.~V.}\ \bibnamefont
  {Braguta}}, \bibinfo {author} {\bibfnamefont {Andrey}\ \bibnamefont {Kotov}},
  \bibinfo {author} {\bibfnamefont {Artem}\ \bibnamefont {Roenko}}, \ and\
  \bibinfo {author} {\bibfnamefont {Dmitry}\ \bibnamefont {Sychev}},\
  }\bibfield  {title} {\enquote {\bibinfo {title} {{Thermal phase transitions
  in rotating QCD with dynamical quarks}},}\ }\href {\doibase
  10.22323/1.430.0190} {\bibfield  {journal} {\bibinfo  {journal} {PoS}\
  }\textbf {\bibinfo {volume} {LATTICE2022}},\ \bibinfo {pages} {190} (\bibinfo
  {year} {2023}{\natexlab{a}})},\ \Eprint {http://arxiv.org/abs/2212.03224}
  {arXiv:2212.03224 [hep-lat]} \BibitemShut {NoStop}%
\bibitem [{\citenamefont {Braguta}\ \emph
  {et~al.}(2023{\natexlab{b}})\citenamefont {Braguta}, \citenamefont {Kudrov},
  \citenamefont {Roenko}, \citenamefont {Sychev},\ and\ \citenamefont
  {Chernodub}}]{Braguta:2023kwl}%
  \BibitemOpen
  \bibfield  {author} {\bibinfo {author} {\bibfnamefont {V.~V.}\ \bibnamefont
  {Braguta}}, \bibinfo {author} {\bibfnamefont {I.~E.}\ \bibnamefont {Kudrov}},
  \bibinfo {author} {\bibfnamefont {A.~A.}\ \bibnamefont {Roenko}}, \bibinfo
  {author} {\bibfnamefont {D.~A.}\ \bibnamefont {Sychev}}, \ and\ \bibinfo
  {author} {\bibfnamefont {M.~N.}\ \bibnamefont {Chernodub}},\ }\bibfield
  {title} {\enquote {\bibinfo {title} {{Lattice Study of the Equation of State
  of a Rotating Gluon Plasma}},}\ }\href {\doibase 10.1134/S0021364023600830}
  {\bibfield  {journal} {\bibinfo  {journal} {JETP Lett.}\ }\textbf {\bibinfo
  {volume} {117}},\ \bibinfo {pages} {639--644} (\bibinfo {year}
  {2023}{\natexlab{b}})}\BibitemShut {NoStop}%
\bibitem [{\citenamefont {Braguta}\ \emph
  {et~al.}(2024{\natexlab{a}})\citenamefont {Braguta}, \citenamefont
  {Chernodub}, \citenamefont {Roenko},\ and\ \citenamefont
  {Sychev}}]{Braguta:2023yjn}%
  \BibitemOpen
  \bibfield  {author} {\bibinfo {author} {\bibfnamefont {Victor~V.}\
  \bibnamefont {Braguta}}, \bibinfo {author} {\bibfnamefont {Maxim~N.}\
  \bibnamefont {Chernodub}}, \bibinfo {author} {\bibfnamefont {Artem~A.}\
  \bibnamefont {Roenko}}, \ and\ \bibinfo {author} {\bibfnamefont {Dmitrii~A.}\
  \bibnamefont {Sychev}},\ }\bibfield  {title} {\enquote {\bibinfo {title}
  {{Negative moment of inertia and rotational instability of gluon plasma}},}\
  }\href {\doibase 10.1016/j.physletb.2024.138604} {\bibfield  {journal}
  {\bibinfo  {journal} {Phys. Lett. B}\ }\textbf {\bibinfo {volume} {852}},\
  \bibinfo {pages} {138604} (\bibinfo {year} {2024}{\natexlab{a}})},\ \Eprint
  {http://arxiv.org/abs/2303.03147} {arXiv:2303.03147 [hep-lat]} \BibitemShut
  {NoStop}%
\bibitem [{\citenamefont {Yang}\ and\ \citenamefont
  {Huang}(2023)}]{Yang:2023vsw}%
  \BibitemOpen
  \bibfield  {author} {\bibinfo {author} {\bibfnamefont {Ji-Chong}\
  \bibnamefont {Yang}}\ and\ \bibinfo {author} {\bibfnamefont {Xu-Guang}\
  \bibnamefont {Huang}},\ }\bibfield  {title} {\enquote {\bibinfo {title} {{QCD
  on Rotating Lattice with Staggered Fermions}},}\ }\href@noop {} {\  (\bibinfo
  {year} {2023})},\ \Eprint {http://arxiv.org/abs/2307.05755} {arXiv:2307.05755
  [hep-lat]} \BibitemShut {NoStop}%
\bibitem [{\citenamefont {Braguta}\ \emph
  {et~al.}(2024{\natexlab{b}})\citenamefont {Braguta}, \citenamefont
  {Chernodub},\ and\ \citenamefont {Roenko}}]{Braguta:2023iyx}%
  \BibitemOpen
  \bibfield  {author} {\bibinfo {author} {\bibfnamefont {Victor~V.}\
  \bibnamefont {Braguta}}, \bibinfo {author} {\bibfnamefont {Maxim~N.}\
  \bibnamefont {Chernodub}}, \ and\ \bibinfo {author} {\bibfnamefont
  {Artem~A.}\ \bibnamefont {Roenko}},\ }\bibfield  {title} {\enquote {\bibinfo
  {title} {{New mixed inhomogeneous phase in vortical gluon plasma:
  First-principle results from rotating SU(3) lattice gauge theory}},}\ }\href
  {\doibase 10.1016/j.physletb.2024.138783} {\bibfield  {journal} {\bibinfo
  {journal} {Phys. Lett. B}\ }\textbf {\bibinfo {volume} {855}},\ \bibinfo
  {pages} {138783} (\bibinfo {year} {2024}{\natexlab{b}})},\ \Eprint
  {http://arxiv.org/abs/2312.13994} {arXiv:2312.13994 [hep-lat]} \BibitemShut
  {NoStop}%
\bibitem [{\citenamefont {Braguta}\ \emph
  {et~al.}(2024{\natexlab{c}})\citenamefont {Braguta}, \citenamefont
  {Chernodub}, \citenamefont {Kudrov}, \citenamefont {Roenko},\ and\
  \citenamefont {Sychev}}]{Braguta:2023tqz}%
  \BibitemOpen
  \bibfield  {author} {\bibinfo {author} {\bibfnamefont {Victor~V.}\
  \bibnamefont {Braguta}}, \bibinfo {author} {\bibfnamefont {Maxim~N.}\
  \bibnamefont {Chernodub}}, \bibinfo {author} {\bibfnamefont {Ilya~E.}\
  \bibnamefont {Kudrov}}, \bibinfo {author} {\bibfnamefont {Artem~A.}\
  \bibnamefont {Roenko}}, \ and\ \bibinfo {author} {\bibfnamefont {Dmitrii~A.}\
  \bibnamefont {Sychev}},\ }\bibfield  {title} {\enquote {\bibinfo {title}
  {{Negative Barnett effect, negative moment of inertia of the gluon plasma,
  and thermal evaporation of the chromomagnetic condensate}},}\ }\href
  {\doibase 10.1103/PhysRevD.110.014511} {\bibfield  {journal} {\bibinfo
  {journal} {Phys. Rev. D}\ }\textbf {\bibinfo {volume} {110}},\ \bibinfo
  {pages} {014511} (\bibinfo {year} {2024}{\natexlab{c}})},\ \Eprint
  {http://arxiv.org/abs/2310.16036} {arXiv:2310.16036 [hep-ph]} \BibitemShut
  {NoStop}%
\bibitem [{\citenamefont {Gell-Mann}\ and\ \citenamefont
  {Lévy}(1960)}]{GellMann1960}%
  \BibitemOpen
  \bibfield  {author} {\bibinfo {author} {\bibfnamefont {M.}~\bibnamefont
  {Gell-Mann}}\ and\ \bibinfo {author} {\bibfnamefont {M.}~\bibnamefont
  {Lévy}},\ }\bibfield  {title} {\enquote {\bibinfo {title} {The axial vector
  current in beta decay},}\ }\href {\doibase 10.1007/bf02859738} {\bibfield
  {journal} {\bibinfo  {journal} {Il Nuovo Cimento}\ }\textbf {\bibinfo
  {volume} {16}},\ \bibinfo {pages} {705–726} (\bibinfo {year}
  {1960})}\BibitemShut {NoStop}%
\bibitem [{\citenamefont {Dumitru}\ and\ \citenamefont
  {Pisarski}(2001)}]{Dumitru:2000in}%
  \BibitemOpen
  \bibfield  {author} {\bibinfo {author} {\bibfnamefont {Adrian}\ \bibnamefont
  {Dumitru}}\ and\ \bibinfo {author} {\bibfnamefont {Robert~D.}\ \bibnamefont
  {Pisarski}},\ }\bibfield  {title} {\enquote {\bibinfo {title}
  {{Event-by-event fluctuations from decay of a Polyakov loop condensate}},}\
  }\href {\doibase 10.1016/S0370-2693(01)00286-6} {\bibfield  {journal}
  {\bibinfo  {journal} {Phys. Lett. B}\ }\textbf {\bibinfo {volume} {504}},\
  \bibinfo {pages} {282--290} (\bibinfo {year} {2001})},\ \Eprint
  {http://arxiv.org/abs/hep-ph/0010083} {arXiv:hep-ph/0010083} \BibitemShut
  {NoStop}%
\bibitem [{\citenamefont {Scavenius}\ \emph
  {et~al.}(2001{\natexlab{a}})\citenamefont {Scavenius}, \citenamefont
  {Dumitru},\ and\ \citenamefont {Jackson}}]{Scavenius:2001pa}%
  \BibitemOpen
  \bibfield  {author} {\bibinfo {author} {\bibfnamefont {O.}~\bibnamefont
  {Scavenius}}, \bibinfo {author} {\bibfnamefont {A.}~\bibnamefont {Dumitru}},
  \ and\ \bibinfo {author} {\bibfnamefont {A.~D.}\ \bibnamefont {Jackson}},\
  }\bibfield  {title} {\enquote {\bibinfo {title} {{Explosive decomposition in
  ultrarelativistic heavy ion collision}},}\ }\href {\doibase
  10.1103/PhysRevLett.87.182302} {\bibfield  {journal} {\bibinfo  {journal}
  {Phys. Rev. Lett.}\ }\textbf {\bibinfo {volume} {87}},\ \bibinfo {pages}
  {182302} (\bibinfo {year} {2001}{\natexlab{a}})},\ \Eprint
  {http://arxiv.org/abs/hep-ph/0103219} {arXiv:hep-ph/0103219} \BibitemShut
  {NoStop}%
\bibitem [{\citenamefont {Scavenius}\ \emph {et~al.}(2002)\citenamefont
  {Scavenius}, \citenamefont {Dumitru},\ and\ \citenamefont
  {Lenaghan}}]{Scavenius:2002ru}%
  \BibitemOpen
  \bibfield  {author} {\bibinfo {author} {\bibfnamefont {O.}~\bibnamefont
  {Scavenius}}, \bibinfo {author} {\bibfnamefont {A.}~\bibnamefont {Dumitru}},
  \ and\ \bibinfo {author} {\bibfnamefont {J.~T.}\ \bibnamefont {Lenaghan}},\
  }\bibfield  {title} {\enquote {\bibinfo {title} {{The $K/\pi$ ratio from
  condensed Polyakov loops}},}\ }\href {\doibase 10.1103/PhysRevC.66.034903}
  {\bibfield  {journal} {\bibinfo  {journal} {Phys. Rev. C}\ }\textbf {\bibinfo
  {volume} {66}},\ \bibinfo {pages} {034903} (\bibinfo {year} {2002})},\
  \Eprint {http://arxiv.org/abs/hep-ph/0201079} {arXiv:hep-ph/0201079}
  \BibitemShut {NoStop}%
\bibitem [{\citenamefont {Megias}\ \emph {et~al.}(2006)\citenamefont {Megias},
  \citenamefont {Ruiz~Arriola},\ and\ \citenamefont {Salcedo}}]{Megias:2004hj}%
  \BibitemOpen
  \bibfield  {author} {\bibinfo {author} {\bibfnamefont {E.}~\bibnamefont
  {Megias}}, \bibinfo {author} {\bibfnamefont {E.}~\bibnamefont
  {Ruiz~Arriola}}, \ and\ \bibinfo {author} {\bibfnamefont {L.~L.}\
  \bibnamefont {Salcedo}},\ }\bibfield  {title} {\enquote {\bibinfo {title}
  {{Polyakov loop in chiral quark models at finite temperature}},}\ }\href
  {\doibase 10.1103/PhysRevD.74.065005} {\bibfield  {journal} {\bibinfo
  {journal} {Phys. Rev. D}\ }\textbf {\bibinfo {volume} {74}},\ \bibinfo
  {pages} {065005} (\bibinfo {year} {2006})},\ \Eprint
  {http://arxiv.org/abs/hep-ph/0412308} {arXiv:hep-ph/0412308} \BibitemShut
  {NoStop}%
\bibitem [{\citenamefont {Schaefer}\ \emph {et~al.}(2007)\citenamefont
  {Schaefer}, \citenamefont {Pawlowski},\ and\ \citenamefont
  {Wambach}}]{Schaefer:2007pw}%
  \BibitemOpen
  \bibfield  {author} {\bibinfo {author} {\bibfnamefont {Bernd-Jochen}\
  \bibnamefont {Schaefer}}, \bibinfo {author} {\bibfnamefont {Jan~M.}\
  \bibnamefont {Pawlowski}}, \ and\ \bibinfo {author} {\bibfnamefont {Jochen}\
  \bibnamefont {Wambach}},\ }\bibfield  {title} {\enquote {\bibinfo {title}
  {{The Phase Structure of the Polyakov--Quark-Meson Model}},}\ }\href
  {\doibase 10.1103/PhysRevD.76.074023} {\bibfield  {journal} {\bibinfo
  {journal} {Phys. Rev. D}\ }\textbf {\bibinfo {volume} {76}},\ \bibinfo
  {pages} {074023} (\bibinfo {year} {2007})},\ \Eprint
  {http://arxiv.org/abs/0704.3234} {arXiv:0704.3234 [hep-ph]} \BibitemShut
  {NoStop}%
\bibitem [{\citenamefont {Chen}\ \emph {et~al.}(2016)\citenamefont {Chen},
  \citenamefont {Fukushima}, \citenamefont {Huang},\ and\ \citenamefont
  {Mameda}}]{Chen:2015hfc}%
  \BibitemOpen
  \bibfield  {author} {\bibinfo {author} {\bibfnamefont {Hao-Lei}\ \bibnamefont
  {Chen}}, \bibinfo {author} {\bibfnamefont {Kenji}\ \bibnamefont {Fukushima}},
  \bibinfo {author} {\bibfnamefont {Xu-Guang}\ \bibnamefont {Huang}}, \ and\
  \bibinfo {author} {\bibfnamefont {Kazuya}\ \bibnamefont {Mameda}},\
  }\bibfield  {title} {\enquote {\bibinfo {title} {{Analogy between rotation
  and density for Dirac fermions in a magnetic field}},}\ }\href {\doibase
  10.1103/PhysRevD.93.104052} {\bibfield  {journal} {\bibinfo  {journal} {Phys.
  Rev. D}\ }\textbf {\bibinfo {volume} {93}},\ \bibinfo {pages} {104052}
  (\bibinfo {year} {2016})},\ \Eprint {http://arxiv.org/abs/1512.08974}
  {arXiv:1512.08974 [hep-ph]} \BibitemShut {NoStop}%
\bibitem [{\citenamefont {Jiang}\ and\ \citenamefont
  {Liao}(2016)}]{Jiang:2016wvv}%
  \BibitemOpen
  \bibfield  {author} {\bibinfo {author} {\bibfnamefont {Yin}\ \bibnamefont
  {Jiang}}\ and\ \bibinfo {author} {\bibfnamefont {Jinfeng}\ \bibnamefont
  {Liao}},\ }\bibfield  {title} {\enquote {\bibinfo {title} {{Pairing Phase
  Transitions of Matter under Rotation}},}\ }\href {\doibase
  10.1103/PhysRevLett.117.192302} {\bibfield  {journal} {\bibinfo  {journal}
  {Phys. Rev. Lett.}\ }\textbf {\bibinfo {volume} {117}},\ \bibinfo {pages}
  {192302} (\bibinfo {year} {2016})},\ \Eprint
  {http://arxiv.org/abs/1606.03808} {arXiv:1606.03808 [hep-ph]} \BibitemShut
  {NoStop}%
\bibitem [{\citenamefont {Chernodub}\ and\ \citenamefont
  {Gongyo}(2017{\natexlab{a}})}]{Chernodub:2017ref}%
  \BibitemOpen
  \bibfield  {author} {\bibinfo {author} {\bibfnamefont {M.~N.}\ \bibnamefont
  {Chernodub}}\ and\ \bibinfo {author} {\bibfnamefont {Shinya}\ \bibnamefont
  {Gongyo}},\ }\bibfield  {title} {\enquote {\bibinfo {title} {{Effects of
  rotation and boundaries on chiral symmetry breaking of relativistic
  fermions}},}\ }\href {\doibase 10.1103/PhysRevD.95.096006} {\bibfield
  {journal} {\bibinfo  {journal} {Phys. Rev. D}\ }\textbf {\bibinfo {volume}
  {95}},\ \bibinfo {pages} {096006} (\bibinfo {year} {2017}{\natexlab{a}})},\
  \Eprint {http://arxiv.org/abs/1702.08266} {arXiv:1702.08266 [hep-th]}
  \BibitemShut {NoStop}%
\bibitem [{\citenamefont {Wang}\ \emph {et~al.}(2019)\citenamefont {Wang},
  \citenamefont {Wei}, \citenamefont {Li},\ and\ \citenamefont
  {Huang}}]{Wang:2018sur}%
  \BibitemOpen
  \bibfield  {author} {\bibinfo {author} {\bibfnamefont {Xinyang}\ \bibnamefont
  {Wang}}, \bibinfo {author} {\bibfnamefont {Minghua}\ \bibnamefont {Wei}},
  \bibinfo {author} {\bibfnamefont {Zhibin}\ \bibnamefont {Li}}, \ and\
  \bibinfo {author} {\bibfnamefont {Mei}\ \bibnamefont {Huang}},\ }\bibfield
  {title} {\enquote {\bibinfo {title} {{Quark matter under rotation in the NJL
  model with vector interaction}},}\ }\href {\doibase
  10.1103/PhysRevD.99.016018} {\bibfield  {journal} {\bibinfo  {journal} {Phys.
  Rev. D}\ }\textbf {\bibinfo {volume} {99}},\ \bibinfo {pages} {016018}
  (\bibinfo {year} {2019})},\ \Eprint {http://arxiv.org/abs/1808.01931}
  {arXiv:1808.01931 [hep-ph]} \BibitemShut {NoStop}%
\bibitem [{\citenamefont {Chen}\ \emph {et~al.}(2021)\citenamefont {Chen},
  \citenamefont {Zhang}, \citenamefont {Li}, \citenamefont {Hou},\ and\
  \citenamefont {Huang}}]{Chen:2020ath}%
  \BibitemOpen
  \bibfield  {author} {\bibinfo {author} {\bibfnamefont {Xun}\ \bibnamefont
  {Chen}}, \bibinfo {author} {\bibfnamefont {Lin}\ \bibnamefont {Zhang}},
  \bibinfo {author} {\bibfnamefont {Danning}\ \bibnamefont {Li}}, \bibinfo
  {author} {\bibfnamefont {Defu}\ \bibnamefont {Hou}}, \ and\ \bibinfo {author}
  {\bibfnamefont {Mei}\ \bibnamefont {Huang}},\ }\bibfield  {title} {\enquote
  {\bibinfo {title} {{Gluodynamics and deconfinement phase transition under
  rotation from holography}},}\ }\href {\doibase 10.1007/JHEP07(2021)132}
  {\bibfield  {journal} {\bibinfo  {journal} {JHEP}\ }\textbf {\bibinfo
  {volume} {07}},\ \bibinfo {pages} {132} (\bibinfo {year} {2021})},\ \Eprint
  {http://arxiv.org/abs/2010.14478} {arXiv:2010.14478 [hep-ph]} \BibitemShut
  {NoStop}%
\bibitem [{\citenamefont {Mehr}\ and\ \citenamefont
  {Taghinavaz}(2023)}]{Mehr:2022tfq}%
  \BibitemOpen
  \bibfield  {author} {\bibinfo {author} {\bibfnamefont {S.~M. A.~Tabatabaee}\
  \bibnamefont {Mehr}}\ and\ \bibinfo {author} {\bibfnamefont {F.}~\bibnamefont
  {Taghinavaz}},\ }\bibfield  {title} {\enquote {\bibinfo {title} {{Chiral
  phase transition of a dense, magnetized and rotating quark matter}},}\ }\href
  {\doibase 10.1016/j.aop.2023.169357} {\bibfield  {journal} {\bibinfo
  {journal} {Annals Phys.}\ }\textbf {\bibinfo {volume} {454}},\ \bibinfo
  {pages} {169357} (\bibinfo {year} {2023})},\ \Eprint
  {http://arxiv.org/abs/2201.05398} {arXiv:2201.05398 [hep-ph]} \BibitemShut
  {NoStop}%
\bibitem [{\citenamefont {Zhao}\ \emph {et~al.}(2023)\citenamefont {Zhao},
  \citenamefont {He}, \citenamefont {Hou}, \citenamefont {Li},\ and\
  \citenamefont {Li}}]{Zhao:2022uxc}%
  \BibitemOpen
  \bibfield  {author} {\bibinfo {author} {\bibfnamefont {Yan-Qing}\
  \bibnamefont {Zhao}}, \bibinfo {author} {\bibfnamefont {Song}\ \bibnamefont
  {He}}, \bibinfo {author} {\bibfnamefont {Defu}\ \bibnamefont {Hou}}, \bibinfo
  {author} {\bibfnamefont {Li}~\bibnamefont {Li}}, \ and\ \bibinfo {author}
  {\bibfnamefont {Zhibin}\ \bibnamefont {Li}},\ }\bibfield  {title} {\enquote
  {\bibinfo {title} {{Phase diagram of holographic thermal dense QCD matter
  with rotation}},}\ }\href {\doibase 10.1007/JHEP04(2023)115} {\bibfield
  {journal} {\bibinfo  {journal} {JHEP}\ }\textbf {\bibinfo {volume} {04}},\
  \bibinfo {pages} {115} (\bibinfo {year} {2023})},\ \Eprint
  {http://arxiv.org/abs/2212.14662} {arXiv:2212.14662 [hep-ph]} \BibitemShut
  {NoStop}%
\bibitem [{\citenamefont {Chen}\ \emph {et~al.}(2023)\citenamefont {Chen},
  \citenamefont {Zhu},\ and\ \citenamefont {Huang}}]{Chen:2023cjt}%
  \BibitemOpen
  \bibfield  {author} {\bibinfo {author} {\bibfnamefont {Hao-Lei}\ \bibnamefont
  {Chen}}, \bibinfo {author} {\bibfnamefont {Zhi-Bin}\ \bibnamefont {Zhu}}, \
  and\ \bibinfo {author} {\bibfnamefont {Xu-Guang}\ \bibnamefont {Huang}},\
  }\bibfield  {title} {\enquote {\bibinfo {title} {{Quark-meson model under
  rotation: A functional renormalization group study}},}\ }\href {\doibase
  10.1103/PhysRevD.108.054006} {\bibfield  {journal} {\bibinfo  {journal}
  {Phys. Rev. D}\ }\textbf {\bibinfo {volume} {108}},\ \bibinfo {pages}
  {054006} (\bibinfo {year} {2023})},\ \Eprint
  {http://arxiv.org/abs/2306.08362} {arXiv:2306.08362 [hep-ph]} \BibitemShut
  {NoStop}%
\bibitem [{\citenamefont {Sun}\ \emph {et~al.}(2023{\natexlab{a}})\citenamefont
  {Sun}, \citenamefont {Li}, \citenamefont {Wen}, \citenamefont {Huang},\ and\
  \citenamefont {Xie}}]{Sun:2023yux}%
  \BibitemOpen
  \bibfield  {author} {\bibinfo {author} {\bibfnamefont {Fei}\ \bibnamefont
  {Sun}}, \bibinfo {author} {\bibfnamefont {Shuang}\ \bibnamefont {Li}},
  \bibinfo {author} {\bibfnamefont {Rui}\ \bibnamefont {Wen}}, \bibinfo
  {author} {\bibfnamefont {Anping}\ \bibnamefont {Huang}}, \ and\ \bibinfo
  {author} {\bibfnamefont {Wei}\ \bibnamefont {Xie}},\ }\bibfield  {title}
  {\enquote {\bibinfo {title} {{The rotation effect on the thermodynamics of
  the QCD matter}},}\ }\href@noop {} {\  (\bibinfo {year}
  {2023}{\natexlab{a}})},\ \Eprint {http://arxiv.org/abs/2310.18942}
  {arXiv:2310.18942 [hep-ph]} \BibitemShut {NoStop}%
\bibitem [{\citenamefont {Barnett}(1915)}]{Barnett:1915uqc}%
  \BibitemOpen
  \bibfield  {author} {\bibinfo {author} {\bibfnamefont {S.~J.}\ \bibnamefont
  {Barnett}},\ }\bibfield  {title} {\enquote {\bibinfo {title} {{Magnetization
  by Rotation}},}\ }\href {\doibase 10.1103/PhysRev.6.239} {\bibfield
  {journal} {\bibinfo  {journal} {Phys. Rev.}\ }\textbf {\bibinfo {volume}
  {6}},\ \bibinfo {pages} {239} (\bibinfo {year} {1915})}\BibitemShut {NoStop}%
\bibitem [{\citenamefont {Sun}\ \emph {et~al.}(2024)\citenamefont {Sun},
  \citenamefont {Shao}, \citenamefont {Wen}, \citenamefont {Xu},\ and\
  \citenamefont {Huang}}]{Sun:2024anu}%
  \BibitemOpen
  \bibfield  {author} {\bibinfo {author} {\bibfnamefont {Fei}\ \bibnamefont
  {Sun}}, \bibinfo {author} {\bibfnamefont {Jingdong}\ \bibnamefont {Shao}},
  \bibinfo {author} {\bibfnamefont {Rui}\ \bibnamefont {Wen}}, \bibinfo
  {author} {\bibfnamefont {Kun}\ \bibnamefont {Xu}}, \ and\ \bibinfo {author}
  {\bibfnamefont {Mei}\ \bibnamefont {Huang}},\ }\bibfield  {title} {\enquote
  {\bibinfo {title} {{Chiral phase transition and spin alignment of vector
  mesons in the polarized-Polyakov-loop Nambu\textendash{}Jona-Lasinio model
  under rotation}},}\ }\href {\doibase 10.1103/PhysRevD.109.116017} {\bibfield
  {journal} {\bibinfo  {journal} {Phys. Rev. D}\ }\textbf {\bibinfo {volume}
  {109}},\ \bibinfo {pages} {116017} (\bibinfo {year} {2024})},\ \Eprint
  {http://arxiv.org/abs/2402.16595} {arXiv:2402.16595 [hep-ph]} \BibitemShut
  {NoStop}%
\bibitem [{\citenamefont {Fukushima}(2004)}]{Fukushima:2003fw}%
  \BibitemOpen
  \bibfield  {author} {\bibinfo {author} {\bibfnamefont {Kenji}\ \bibnamefont
  {Fukushima}},\ }\bibfield  {title} {\enquote {\bibinfo {title} {{Chiral
  effective model with the Polyakov loop}},}\ }\href {\doibase
  10.1016/j.physletb.2004.04.027} {\bibfield  {journal} {\bibinfo  {journal}
  {Phys. Lett. B}\ }\textbf {\bibinfo {volume} {591}},\ \bibinfo {pages}
  {277--284} (\bibinfo {year} {2004})},\ \Eprint
  {http://arxiv.org/abs/hep-ph/0310121} {arXiv:hep-ph/0310121} \BibitemShut
  {NoStop}%
\bibitem [{\citenamefont {Roessner}\ \emph {et~al.}(2007)\citenamefont
  {Roessner}, \citenamefont {Ratti},\ and\ \citenamefont
  {Weise}}]{Roessner:2006xn}%
  \BibitemOpen
  \bibfield  {author} {\bibinfo {author} {\bibfnamefont {Simon}\ \bibnamefont
  {Roessner}}, \bibinfo {author} {\bibfnamefont {Claudia}\ \bibnamefont
  {Ratti}}, \ and\ \bibinfo {author} {\bibfnamefont {W.}~\bibnamefont
  {Weise}},\ }\bibfield  {title} {\enquote {\bibinfo {title} {{Polyakov loop,
  diquarks and the two-flavour phase diagram}},}\ }\href {\doibase
  10.1103/PhysRevD.75.034007} {\bibfield  {journal} {\bibinfo  {journal} {Phys.
  Rev. D}\ }\textbf {\bibinfo {volume} {75}},\ \bibinfo {pages} {034007}
  (\bibinfo {year} {2007})},\ \Eprint {http://arxiv.org/abs/hep-ph/0609281}
  {arXiv:hep-ph/0609281} \BibitemShut {NoStop}%
\bibitem [{\citenamefont {Farias}\ \emph {et~al.}(2017)\citenamefont {Farias},
  \citenamefont {Timoteo}, \citenamefont {Avancini}, \citenamefont {Pinto},\
  and\ \citenamefont {Krein}}]{Farias:2016gmy}%
  \BibitemOpen
  \bibfield  {author} {\bibinfo {author} {\bibfnamefont {R.~L.~S.}\
  \bibnamefont {Farias}}, \bibinfo {author} {\bibfnamefont {V.~S.}\
  \bibnamefont {Timoteo}}, \bibinfo {author} {\bibfnamefont {S.~S.}\
  \bibnamefont {Avancini}}, \bibinfo {author} {\bibfnamefont {M.~B.}\
  \bibnamefont {Pinto}}, \ and\ \bibinfo {author} {\bibfnamefont
  {G.}~\bibnamefont {Krein}},\ }\bibfield  {title} {\enquote {\bibinfo {title}
  {{Thermo-magnetic effects in quark matter: Nambu--Jona-Lasinio model
  constrained by lattice QCD}},}\ }\href {\doibase 10.1140/epja/i2017-12320-8}
  {\bibfield  {journal} {\bibinfo  {journal} {Eur. Phys. J. A}\ }\textbf
  {\bibinfo {volume} {53}},\ \bibinfo {pages} {101} (\bibinfo {year} {2017})},\
  \Eprint {http://arxiv.org/abs/1603.03847} {arXiv:1603.03847 [hep-ph]}
  \BibitemShut {NoStop}%
\bibitem [{\citenamefont {Nambu}\ and\ \citenamefont
  {Jona-Lasinio}(1961{\natexlab{a}})}]{Nambu:1961tp}%
  \BibitemOpen
  \bibfield  {author} {\bibinfo {author} {\bibfnamefont {Yoichiro}\
  \bibnamefont {Nambu}}\ and\ \bibinfo {author} {\bibfnamefont
  {G.}~\bibnamefont {Jona-Lasinio}},\ }\bibfield  {title} {\enquote {\bibinfo
  {title} {{Dynamical Model of Elementary Particles Based on an Analogy with
  Superconductivity. 1.}}}\ }\href {\doibase 10.1103/PhysRev.122.345}
  {\bibfield  {journal} {\bibinfo  {journal} {Phys. Rev.}\ }\textbf {\bibinfo
  {volume} {122}},\ \bibinfo {pages} {345--358} (\bibinfo {year}
  {1961}{\natexlab{a}})}\BibitemShut {NoStop}%
\bibitem [{\citenamefont {Nambu}\ and\ \citenamefont
  {Jona-Lasinio}(1961{\natexlab{b}})}]{Nambu:1961fr}%
  \BibitemOpen
  \bibfield  {author} {\bibinfo {author} {\bibfnamefont {Yoichiro}\
  \bibnamefont {Nambu}}\ and\ \bibinfo {author} {\bibfnamefont
  {G.}~\bibnamefont {Jona-Lasinio}},\ }\bibfield  {title} {\enquote {\bibinfo
  {title} {{Dynamical model of elementary particles based on an analogy with
  superconductivity. II.}}}\ }\href {\doibase 10.1103/PhysRev.124.246}
  {\bibfield  {journal} {\bibinfo  {journal} {Phys. Rev.}\ }\textbf {\bibinfo
  {volume} {124}},\ \bibinfo {pages} {246--254} (\bibinfo {year}
  {1961}{\natexlab{b}})}\BibitemShut {NoStop}%
\bibitem [{\citenamefont {Fukushima}\ and\ \citenamefont
  {Skokov}(2017)}]{Fukushima:2017csk}%
  \BibitemOpen
  \bibfield  {author} {\bibinfo {author} {\bibfnamefont {Kenji}\ \bibnamefont
  {Fukushima}}\ and\ \bibinfo {author} {\bibfnamefont {Vladimir}\ \bibnamefont
  {Skokov}},\ }\bibfield  {title} {\enquote {\bibinfo {title} {{Polyakov loop
  modeling for hot QCD}},}\ }\href {\doibase 10.1016/j.ppnp.2017.05.002}
  {\bibfield  {journal} {\bibinfo  {journal} {Prog. Part. Nucl. Phys.}\
  }\textbf {\bibinfo {volume} {96}},\ \bibinfo {pages} {154--199} (\bibinfo
  {year} {2017})},\ \Eprint {http://arxiv.org/abs/1705.00718} {arXiv:1705.00718
  [hep-ph]} \BibitemShut {NoStop}%
\bibitem [{\citenamefont {Sun}\ \emph {et~al.}(2023{\natexlab{b}})\citenamefont
  {Sun}, \citenamefont {Xu},\ and\ \citenamefont {Huang}}]{Sun:2023kuu}%
  \BibitemOpen
  \bibfield  {author} {\bibinfo {author} {\bibfnamefont {Fei}\ \bibnamefont
  {Sun}}, \bibinfo {author} {\bibfnamefont {Kun}\ \bibnamefont {Xu}}, \ and\
  \bibinfo {author} {\bibfnamefont {Mei}\ \bibnamefont {Huang}},\ }\bibfield
  {title} {\enquote {\bibinfo {title} {{Splitting of chiral and deconfinement
  phase transitions induced by rotation}},}\ }\href {\doibase
  10.1103/PhysRevD.108.096007} {\bibfield  {journal} {\bibinfo  {journal}
  {Phys. Rev. D}\ }\textbf {\bibinfo {volume} {108}},\ \bibinfo {pages}
  {096007} (\bibinfo {year} {2023}{\natexlab{b}})},\ \Eprint
  {http://arxiv.org/abs/2307.14402} {arXiv:2307.14402 [hep-ph]} \BibitemShut
  {NoStop}%
\bibitem [{\citenamefont {Davies}\ \emph {et~al.}(1996)\citenamefont {Davies},
  \citenamefont {Dray},\ and\ \citenamefont {Manogue}}]{Davies:1996ks}%
  \BibitemOpen
  \bibfield  {author} {\bibinfo {author} {\bibfnamefont {Paul C.~W.}\
  \bibnamefont {Davies}}, \bibinfo {author} {\bibfnamefont {Tevian}\
  \bibnamefont {Dray}}, \ and\ \bibinfo {author} {\bibfnamefont {Corinne~A.}\
  \bibnamefont {Manogue}},\ }\bibfield  {title} {\enquote {\bibinfo {title}
  {{The Rotating quantum vacuum}},}\ }\href {\doibase 10.1103/PhysRevD.53.4382}
  {\bibfield  {journal} {\bibinfo  {journal} {Phys. Rev. D}\ }\textbf {\bibinfo
  {volume} {53}},\ \bibinfo {pages} {4382--4387} (\bibinfo {year} {1996})},\
  \Eprint {http://arxiv.org/abs/gr-qc/9601034} {arXiv:gr-qc/9601034}
  \BibitemShut {NoStop}%
\bibitem [{\citenamefont {Scavenius}\ \emph
  {et~al.}(2001{\natexlab{b}})\citenamefont {Scavenius}, \citenamefont {Mocsy},
  \citenamefont {Mishustin},\ and\ \citenamefont {Rischke}}]{Scavenius:2000qd}%
  \BibitemOpen
  \bibfield  {author} {\bibinfo {author} {\bibfnamefont {O.}~\bibnamefont
  {Scavenius}}, \bibinfo {author} {\bibfnamefont {A.}~\bibnamefont {Mocsy}},
  \bibinfo {author} {\bibfnamefont {I.~N.}\ \bibnamefont {Mishustin}}, \ and\
  \bibinfo {author} {\bibfnamefont {D.~H.}\ \bibnamefont {Rischke}},\
  }\bibfield  {title} {\enquote {\bibinfo {title} {{Chiral phase transition
  within effective models with constituent quarks}},}\ }\href {\doibase
  10.1103/PhysRevC.64.045202} {\bibfield  {journal} {\bibinfo  {journal} {Phys.
  Rev. C}\ }\textbf {\bibinfo {volume} {64}},\ \bibinfo {pages} {045202}
  (\bibinfo {year} {2001}{\natexlab{b}})},\ \Eprint
  {http://arxiv.org/abs/nucl-th/0007030} {arXiv:nucl-th/0007030} \BibitemShut
  {NoStop}%
\bibitem [{\citenamefont {Pisarski}\ and\ \citenamefont
  {Wilczek}(1984)}]{Pisarski:1983ms}%
  \BibitemOpen
  \bibfield  {author} {\bibinfo {author} {\bibfnamefont {Robert~D.}\
  \bibnamefont {Pisarski}}\ and\ \bibinfo {author} {\bibfnamefont {Frank}\
  \bibnamefont {Wilczek}},\ }\bibfield  {title} {\enquote {\bibinfo {title}
  {{Remarks on the Chiral Phase Transition in Chromodynamics}},}\ }\href
  {\doibase 10.1103/PhysRevD.29.338} {\bibfield  {journal} {\bibinfo  {journal}
  {Phys. Rev. D}\ }\textbf {\bibinfo {volume} {29}},\ \bibinfo {pages}
  {338--341} (\bibinfo {year} {1984})}\BibitemShut {NoStop}%
\bibitem [{\citenamefont {Scavenius}\ \emph
  {et~al.}(2001{\natexlab{c}})\citenamefont {Scavenius}, \citenamefont
  {Dumitru}, \citenamefont {Fraga}, \citenamefont {Lenaghan},\ and\
  \citenamefont {Jackson}}]{Scavenius:2000bb}%
  \BibitemOpen
  \bibfield  {author} {\bibinfo {author} {\bibfnamefont {O.}~\bibnamefont
  {Scavenius}}, \bibinfo {author} {\bibfnamefont {A.}~\bibnamefont {Dumitru}},
  \bibinfo {author} {\bibfnamefont {E.~S.}\ \bibnamefont {Fraga}}, \bibinfo
  {author} {\bibfnamefont {J.~T.}\ \bibnamefont {Lenaghan}}, \ and\ \bibinfo
  {author} {\bibfnamefont {A.~D.}\ \bibnamefont {Jackson}},\ }\bibfield
  {title} {\enquote {\bibinfo {title} {{First order chiral phase transition in
  high-energy collisions: Can nucleation prevent spinodal decomposition?}}}\
  }\href {\doibase 10.1103/PhysRevD.63.116003} {\bibfield  {journal} {\bibinfo
  {journal} {Phys. Rev. D}\ }\textbf {\bibinfo {volume} {63}},\ \bibinfo
  {pages} {116003} (\bibinfo {year} {2001}{\natexlab{c}})},\ \Eprint
  {http://arxiv.org/abs/hep-ph/0009171} {arXiv:hep-ph/0009171} \BibitemShut
  {NoStop}%
\bibitem [{\citenamefont {Skokov}\ \emph {et~al.}(2010)\citenamefont {Skokov},
  \citenamefont {Stokic}, \citenamefont {Friman},\ and\ \citenamefont
  {Redlich}}]{Skokov:2010wb}%
  \BibitemOpen
  \bibfield  {author} {\bibinfo {author} {\bibfnamefont {V.}~\bibnamefont
  {Skokov}}, \bibinfo {author} {\bibfnamefont {B.}~\bibnamefont {Stokic}},
  \bibinfo {author} {\bibfnamefont {B.}~\bibnamefont {Friman}}, \ and\ \bibinfo
  {author} {\bibfnamefont {K.}~\bibnamefont {Redlich}},\ }\bibfield  {title}
  {\enquote {\bibinfo {title} {{Meson fluctuations and thermodynamics of the
  Polyakov loop extended quark-meson model}},}\ }\href {\doibase
  10.1103/PhysRevC.82.015206} {\bibfield  {journal} {\bibinfo  {journal} {Phys.
  Rev. C}\ }\textbf {\bibinfo {volume} {82}},\ \bibinfo {pages} {015206}
  (\bibinfo {year} {2010})},\ \Eprint {http://arxiv.org/abs/1004.2665}
  {arXiv:1004.2665 [hep-ph]} \BibitemShut {NoStop}%
\bibitem [{\citenamefont {Roessner}\ \emph {et~al.}(2008)\citenamefont
  {Roessner}, \citenamefont {Hell}, \citenamefont {Ratti},\ and\ \citenamefont
  {Weise}}]{Roessner:2007gha}%
  \BibitemOpen
  \bibfield  {author} {\bibinfo {author} {\bibfnamefont {Simon}\ \bibnamefont
  {Roessner}}, \bibinfo {author} {\bibfnamefont {T.}~\bibnamefont {Hell}},
  \bibinfo {author} {\bibfnamefont {C.}~\bibnamefont {Ratti}}, \ and\ \bibinfo
  {author} {\bibfnamefont {W.}~\bibnamefont {Weise}},\ }\bibfield  {title}
  {\enquote {\bibinfo {title} {{The chiral and deconfinement crossover
  transitions: PNJL model beyond mean field}},}\ }\href {\doibase
  10.1016/j.nuclphysa.2008.10.006} {\bibfield  {journal} {\bibinfo  {journal}
  {Nucl. Phys. A}\ }\textbf {\bibinfo {volume} {814}},\ \bibinfo {pages}
  {118--143} (\bibinfo {year} {2008})},\ \Eprint
  {http://arxiv.org/abs/0712.3152} {arXiv:0712.3152 [hep-ph]} \BibitemShut
  {NoStop}%
\bibitem [{\citenamefont {Chen}\ \emph {et~al.}(2022)\citenamefont {Chen},
  \citenamefont {Fukushima},\ and\ \citenamefont {Shimada}}]{Chen:2022smf}%
  \BibitemOpen
  \bibfield  {author} {\bibinfo {author} {\bibfnamefont {Shi}\ \bibnamefont
  {Chen}}, \bibinfo {author} {\bibfnamefont {Kenji}\ \bibnamefont {Fukushima}},
  \ and\ \bibinfo {author} {\bibfnamefont {Yusuke}\ \bibnamefont {Shimada}},\
  }\bibfield  {title} {\enquote {\bibinfo {title} {{Perturbative Confinement in
  Thermal Yang-Mills Theories Induced by Imaginary Angular Velocity}},}\ }\href
  {\doibase 10.1103/PhysRevLett.129.242002} {\bibfield  {journal} {\bibinfo
  {journal} {Phys. Rev. Lett.}\ }\textbf {\bibinfo {volume} {129}},\ \bibinfo
  {pages} {242002} (\bibinfo {year} {2022})},\ \Eprint
  {http://arxiv.org/abs/2207.12665} {arXiv:2207.12665 [hep-ph]} \BibitemShut
  {NoStop}%
\bibitem [{\citenamefont {Chen}\ \emph {et~al.}(2024)\citenamefont {Chen},
  \citenamefont {Fukushima},\ and\ \citenamefont {Shimada}}]{Chen:2024tkr}%
  \BibitemOpen
  \bibfield  {author} {\bibinfo {author} {\bibfnamefont {Shi}\ \bibnamefont
  {Chen}}, \bibinfo {author} {\bibfnamefont {Kenji}\ \bibnamefont {Fukushima}},
  \ and\ \bibinfo {author} {\bibfnamefont {Yusuke}\ \bibnamefont {Shimada}},\
  }\bibfield  {title} {\enquote {\bibinfo {title} {{Inhomogeneous confinement
  and chiral symmetry breaking induced by imaginary angular velocity}},}\
  }\href@noop {} {\  (\bibinfo {year} {2024})},\ \Eprint
  {http://arxiv.org/abs/2404.00965} {arXiv:2404.00965 [hep-ph]} \BibitemShut
  {NoStop}%
\bibitem [{\citenamefont {Kapusta}\ and\ \citenamefont
  {Gale}(2006)}]{kapusta_gale_2006}%
  \BibitemOpen
  \bibfield  {author} {\bibinfo {author} {\bibfnamefont {Joseph~I.}\
  \bibnamefont {Kapusta}}\ and\ \bibinfo {author} {\bibfnamefont {Charles}\
  \bibnamefont {Gale}},\ }\href {\doibase 10.1017/CBO9780511535130} {\emph
  {\bibinfo {title} {Finite-Temperature Field Theory: Principles and
  Applications}}},\ \bibinfo {edition} {2nd}\ ed.,\ Cambridge Monographs on
  Mathematical Physics\ (\bibinfo  {publisher} {Cambridge University Press},\
  \bibinfo {year} {2006})\BibitemShut {NoStop}%
\bibitem [{\citenamefont {Singha}\ \emph {et~al.}(2024)\citenamefont {Singha},
  \citenamefont {Ambrus},\ and\ \citenamefont {Chernodub}}]{in_preparation}%
  \BibitemOpen
  \bibfield  {author} {\bibinfo {author} {\bibfnamefont {Pracheta}\
  \bibnamefont {Singha}}, \bibinfo {author} {\bibfnamefont {Victor~E.}\
  \bibnamefont {Ambrus}}, \ and\ \bibinfo {author} {\bibfnamefont {M.~N.}\
  \bibnamefont {Chernodub}},\ }\href@noop {} {\bibfield  {journal} {\bibinfo
  {journal} {{in preparation}}\ } (\bibinfo {year} {2024})}\BibitemShut
  {NoStop}%
\bibitem [{\citenamefont {Ratti}\ \emph {et~al.}(2007)\citenamefont {Ratti},
  \citenamefont {Roessner},\ and\ \citenamefont {Weise}}]{Ratti:2007jf}%
  \BibitemOpen
  \bibfield  {author} {\bibinfo {author} {\bibfnamefont {Claudia}\ \bibnamefont
  {Ratti}}, \bibinfo {author} {\bibfnamefont {Simon}\ \bibnamefont {Roessner}},
  \ and\ \bibinfo {author} {\bibfnamefont {Wolfram}\ \bibnamefont {Weise}},\
  }\bibfield  {title} {\enquote {\bibinfo {title} {{Quark number
  susceptibilities: Lattice QCD versus PNJL model}},}\ }\href {\doibase
  10.1016/j.physletb.2007.03.038} {\bibfield  {journal} {\bibinfo  {journal}
  {Phys. Lett. B}\ }\textbf {\bibinfo {volume} {649}},\ \bibinfo {pages}
  {57--60} (\bibinfo {year} {2007})},\ \Eprint
  {http://arxiv.org/abs/hep-ph/0701091} {arXiv:hep-ph/0701091} \BibitemShut
  {NoStop}%
\bibitem [{\citenamefont {Torres-Rincon}\ and\ \citenamefont
  {Aichelin}(2017)}]{Torres-Rincon:2017zbr}%
  \BibitemOpen
  \bibfield  {author} {\bibinfo {author} {\bibfnamefont {Juan~M.}\ \bibnamefont
  {Torres-Rincon}}\ and\ \bibinfo {author} {\bibfnamefont {Joerg}\ \bibnamefont
  {Aichelin}},\ }\bibfield  {title} {\enquote {\bibinfo {title} {{Equation of
  state of a quark-meson mixture in the improved
  Polyakov\textendash{}Nambu\textendash{}Jona-Lasinio model at finite chemical
  potential}},}\ }\href {\doibase 10.1103/PhysRevC.96.045205} {\bibfield
  {journal} {\bibinfo  {journal} {Phys. Rev. C}\ }\textbf {\bibinfo {volume}
  {96}},\ \bibinfo {pages} {045205} (\bibinfo {year} {2017})},\ \Eprint
  {http://arxiv.org/abs/1704.07858} {arXiv:1704.07858 [nucl-th]} \BibitemShut
  {NoStop}%
\bibitem [{\citenamefont {Kashiwa}\ \emph {et~al.}(2008)\citenamefont
  {Kashiwa}, \citenamefont {Kouno}, \citenamefont {Matsuzaki},\ and\
  \citenamefont {Yahiro}}]{Kashiwa:2007hw}%
  \BibitemOpen
  \bibfield  {author} {\bibinfo {author} {\bibfnamefont {Kouji}\ \bibnamefont
  {Kashiwa}}, \bibinfo {author} {\bibfnamefont {Hiroaki}\ \bibnamefont
  {Kouno}}, \bibinfo {author} {\bibfnamefont {Masayuki}\ \bibnamefont
  {Matsuzaki}}, \ and\ \bibinfo {author} {\bibfnamefont {Masanobu}\
  \bibnamefont {Yahiro}},\ }\bibfield  {title} {\enquote {\bibinfo {title}
  {{Critical endpoint in the Polyakov-loop extended NJL model}},}\ }\href
  {\doibase 10.1016/j.physletb.2008.01.075} {\bibfield  {journal} {\bibinfo
  {journal} {Phys. Lett. B}\ }\textbf {\bibinfo {volume} {662}},\ \bibinfo
  {pages} {26--32} (\bibinfo {year} {2008})},\ \Eprint
  {http://arxiv.org/abs/0710.2180} {arXiv:0710.2180 [hep-ph]} \BibitemShut
  {NoStop}%
\bibitem [{\citenamefont {Greensite}(2003)}]{Greensite:2003bk}%
  \BibitemOpen
  \bibfield  {author} {\bibinfo {author} {\bibfnamefont {J.}~\bibnamefont
  {Greensite}},\ }\bibfield  {title} {\enquote {\bibinfo {title} {{The
  Confinement problem in lattice gauge theory}},}\ }\href {\doibase
  10.1016/S0146-6410(03)90012-3} {\bibfield  {journal} {\bibinfo  {journal}
  {Prog. Part. Nucl. Phys.}\ }\textbf {\bibinfo {volume} {51}},\ \bibinfo
  {pages} {1} (\bibinfo {year} {2003})},\ \Eprint
  {http://arxiv.org/abs/hep-lat/0301023} {arXiv:hep-lat/0301023} \BibitemShut
  {NoStop}%
\bibitem [{\citenamefont {Fukushima}\ and\ \citenamefont
  {Sasaki}(2013)}]{Fukushima:2013rx}%
  \BibitemOpen
  \bibfield  {author} {\bibinfo {author} {\bibfnamefont {Kenji}\ \bibnamefont
  {Fukushima}}\ and\ \bibinfo {author} {\bibfnamefont {Chihiro}\ \bibnamefont
  {Sasaki}},\ }\bibfield  {title} {\enquote {\bibinfo {title} {{The phase
  diagram of nuclear and quark matter at high baryon density}},}\ }\href
  {\doibase 10.1016/j.ppnp.2013.05.003} {\bibfield  {journal} {\bibinfo
  {journal} {Prog. Part. Nucl. Phys.}\ }\textbf {\bibinfo {volume} {72}},\
  \bibinfo {pages} {99--154} (\bibinfo {year} {2013})},\ \Eprint
  {http://arxiv.org/abs/1301.6377} {arXiv:1301.6377 [hep-ph]} \BibitemShut
  {NoStop}%
\bibitem [{\citenamefont {Ambru\c{s}}\ and\ \citenamefont
  {Winstanley}(2014)}]{Ambrus:2014uqa}%
  \BibitemOpen
  \bibfield  {author} {\bibinfo {author} {\bibfnamefont {Victor~E.}\
  \bibnamefont {Ambru\c{s}}}\ and\ \bibinfo {author} {\bibfnamefont
  {Elizabeth}\ \bibnamefont {Winstanley}},\ }\bibfield  {title} {\enquote
  {\bibinfo {title} {{Rotating quantum states}},}\ }\href {\doibase
  10.1016/j.physletb.2014.05.031} {\bibfield  {journal} {\bibinfo  {journal}
  {Phys. Lett. B}\ }\textbf {\bibinfo {volume} {734}},\ \bibinfo {pages}
  {296--301} (\bibinfo {year} {2014})},\ \Eprint
  {http://arxiv.org/abs/1401.6388} {arXiv:1401.6388 [hep-th]} \BibitemShut
  {NoStop}%
\bibitem [{\citenamefont {Chernodub}\ and\ \citenamefont
  {Gongyo}(2017{\natexlab{b}})}]{Chernodub:2016kxh}%
  \BibitemOpen
  \bibfield  {author} {\bibinfo {author} {\bibfnamefont {M.~N.}\ \bibnamefont
  {Chernodub}}\ and\ \bibinfo {author} {\bibfnamefont {Shinya}\ \bibnamefont
  {Gongyo}},\ }\bibfield  {title} {\enquote {\bibinfo {title} {{Interacting
  fermions in rotation: chiral symmetry restoration, moment of inertia and
  thermodynamics}},}\ }\href {\doibase 10.1007/JHEP01(2017)136} {\bibfield
  {journal} {\bibinfo  {journal} {JHEP}\ }\textbf {\bibinfo {volume} {01}},\
  \bibinfo {pages} {136} (\bibinfo {year} {2017}{\natexlab{b}})},\ \Eprint
  {http://arxiv.org/abs/1611.02598} {arXiv:1611.02598 [hep-th]} \BibitemShut
  {NoStop}%
\bibitem [{\citenamefont {Landau}\ and\ \citenamefont {Lifshitz}(1982)}]{LL1}%
  \BibitemOpen
  \bibfield  {author} {\bibinfo {author} {\bibfnamefont {L~D}\ \bibnamefont
  {Landau}}\ and\ \bibinfo {author} {\bibfnamefont {E~M}\ \bibnamefont
  {Lifshitz}},\ }\href@noop {} {\emph {\bibinfo {title} {Mechanics}}},\
  \bibinfo {edition} {3rd}\ ed.\ (\bibinfo  {publisher}
  {Butterworth-Heinemann},\ \bibinfo {address} {Oxford, England},\ \bibinfo
  {year} {1982})\BibitemShut {NoStop}%
\bibitem [{\citenamefont {Landau}\ and\ \citenamefont {Lifshitz}(1996)}]{LL5}%
  \BibitemOpen
  \bibfield  {author} {\bibinfo {author} {\bibfnamefont {L.~D.}\ \bibnamefont
  {Landau}}\ and\ \bibinfo {author} {\bibfnamefont {E.~M.}\ \bibnamefont
  {Lifshitz}},\ }\href@noop {} {\emph {\bibinfo {title} {Statistical
  Physics}}},\ \bibinfo {edition} {3rd}\ ed.\ (\bibinfo  {publisher}
  {Butterworth-Heinemann},\ \bibinfo {address} {Oxford, England},\ \bibinfo
  {year} {1996})\BibitemShut {NoStop}%
\bibitem [{\citenamefont {Ambrus}\ and\ \citenamefont
  {Winstanley}(2016)}]{Ambrus:2015lfr}%
  \BibitemOpen
  \bibfield  {author} {\bibinfo {author} {\bibfnamefont {Victor~E.}\
  \bibnamefont {Ambrus}}\ and\ \bibinfo {author} {\bibfnamefont {Elizabeth}\
  \bibnamefont {Winstanley}},\ }\bibfield  {title} {\enquote {\bibinfo {title}
  {{Rotating fermions inside a cylindrical boundary}},}\ }\href {\doibase
  10.1103/PhysRevD.93.104014} {\bibfield  {journal} {\bibinfo  {journal} {Phys.
  Rev. D}\ }\textbf {\bibinfo {volume} {93}},\ \bibinfo {pages} {104014}
  (\bibinfo {year} {2016})},\ \Eprint {http://arxiv.org/abs/1512.05239}
  {arXiv:1512.05239 [hep-th]} \BibitemShut {NoStop}%
\bibitem [{\citenamefont {Hortacsu}\ \emph {et~al.}(1980)\citenamefont
  {Hortacsu}, \citenamefont {Rothe},\ and\ \citenamefont
  {Schroer}}]{Hortacsu:1980kv}%
  \BibitemOpen
  \bibfield  {author} {\bibinfo {author} {\bibfnamefont {M.}~\bibnamefont
  {Hortacsu}}, \bibinfo {author} {\bibfnamefont {K.~D.}\ \bibnamefont {Rothe}},
  \ and\ \bibinfo {author} {\bibfnamefont {B.}~\bibnamefont {Schroer}},\
  }\bibfield  {title} {\enquote {\bibinfo {title} {{Zero Energy Eigenstates for
  the Dirac Boundary Problem}},}\ }\href {\doibase
  10.1016/0550-3213(80)90384-3} {\bibfield  {journal} {\bibinfo  {journal}
  {Nucl. Phys. B}\ }\textbf {\bibinfo {volume} {171}},\ \bibinfo {pages}
  {530--542} (\bibinfo {year} {1980})}\BibitemShut {NoStop}%
\bibitem [{\citenamefont {Chernodub}(2021)}]{Chernodub:2020qah}%
  \BibitemOpen
  \bibfield  {author} {\bibinfo {author} {\bibfnamefont {M.~N.}\ \bibnamefont
  {Chernodub}},\ }\bibfield  {title} {\enquote {\bibinfo {title}
  {{Inhomogeneous confining-deconfining phases in rotating plasmas}},}\ }\href
  {\doibase 10.1103/PhysRevD.103.054027} {\bibfield  {journal} {\bibinfo
  {journal} {Phys. Rev. D}\ }\textbf {\bibinfo {volume} {103}},\ \bibinfo
  {pages} {054027} (\bibinfo {year} {2021})},\ \Eprint
  {http://arxiv.org/abs/2012.04924} {arXiv:2012.04924 [hep-ph]} \BibitemShut
  {NoStop}%
\bibitem [{\citenamefont {Chernodub}\ \emph {et~al.}(2023)\citenamefont
  {Chernodub}, \citenamefont {Goy},\ and\ \citenamefont
  {Molochkov}}]{Chernodub:2022veq}%
  \BibitemOpen
  \bibfield  {author} {\bibinfo {author} {\bibfnamefont {M.~N.}\ \bibnamefont
  {Chernodub}}, \bibinfo {author} {\bibfnamefont {V.~A.}\ \bibnamefont {Goy}},
  \ and\ \bibinfo {author} {\bibfnamefont {A.~V.}\ \bibnamefont {Molochkov}},\
  }\bibfield  {title} {\enquote {\bibinfo {title} {{Inhomogeneity of a rotating
  gluon plasma and the Tolman-Ehrenfest law in imaginary time: Lattice results
  for fast imaginary rotation}},}\ }\href {\doibase
  10.1103/PhysRevD.107.114502} {\bibfield  {journal} {\bibinfo  {journal}
  {Phys. Rev. D}\ }\textbf {\bibinfo {volume} {107}},\ \bibinfo {pages}
  {114502} (\bibinfo {year} {2023})},\ \Eprint
  {http://arxiv.org/abs/2209.15534} {arXiv:2209.15534 [hep-lat]} \BibitemShut
  {NoStop}%
\bibitem [{\citenamefont {Aoki}\ \emph {et~al.}(2006)\citenamefont {Aoki},
  \citenamefont {Fodor}, \citenamefont {Katz},\ and\ \citenamefont
  {Szabo}}]{Aoki:2006br}%
  \BibitemOpen
  \bibfield  {author} {\bibinfo {author} {\bibfnamefont {Y.}~\bibnamefont
  {Aoki}}, \bibinfo {author} {\bibfnamefont {Z.}~\bibnamefont {Fodor}},
  \bibinfo {author} {\bibfnamefont {S.~D.}\ \bibnamefont {Katz}}, \ and\
  \bibinfo {author} {\bibfnamefont {K.~K.}\ \bibnamefont {Szabo}},\ }\bibfield
  {title} {\enquote {\bibinfo {title} {{The QCD transition temperature: Results
  with physical masses in the continuum limit}},}\ }\href {\doibase
  10.1016/j.physletb.2006.10.021} {\bibfield  {journal} {\bibinfo  {journal}
  {Phys. Lett. B}\ }\textbf {\bibinfo {volume} {643}},\ \bibinfo {pages}
  {46--54} (\bibinfo {year} {2006})},\ \Eprint
  {http://arxiv.org/abs/hep-lat/0609068} {arXiv:hep-lat/0609068} \BibitemShut
  {NoStop}%
\bibitem [{\citenamefont {Borsanyi}\ \emph {et~al.}(2010)\citenamefont
  {Borsanyi}, \citenamefont {Fodor}, \citenamefont {Hoelbling}, \citenamefont
  {Katz}, \citenamefont {Krieg}, \citenamefont {Ratti},\ and\ \citenamefont
  {Szabo}}]{Borsanyi:2010bp}%
  \BibitemOpen
  \bibfield  {author} {\bibinfo {author} {\bibfnamefont {Szabolcs}\
  \bibnamefont {Borsanyi}}, \bibinfo {author} {\bibfnamefont {Zoltan}\
  \bibnamefont {Fodor}}, \bibinfo {author} {\bibfnamefont {Christian}\
  \bibnamefont {Hoelbling}}, \bibinfo {author} {\bibfnamefont {Sandor~D}\
  \bibnamefont {Katz}}, \bibinfo {author} {\bibfnamefont {Stefan}\ \bibnamefont
  {Krieg}}, \bibinfo {author} {\bibfnamefont {Claudia}\ \bibnamefont {Ratti}},
  \ and\ \bibinfo {author} {\bibfnamefont {Kalman~K.}\ \bibnamefont {Szabo}}
  (\bibinfo {collaboration} {Wuppertal-Budapest}),\ }\bibfield  {title}
  {\enquote {\bibinfo {title} {{Is there still any $T_c$ mystery in lattice
  QCD? Results with physical masses in the continuum limit III}},}\ }\href
  {\doibase 10.1007/JHEP09(2010)073} {\bibfield  {journal} {\bibinfo  {journal}
  {JHEP}\ }\textbf {\bibinfo {volume} {09}},\ \bibinfo {pages} {073} (\bibinfo
  {year} {2010})},\ \Eprint {http://arxiv.org/abs/1005.3508} {arXiv:1005.3508
  [hep-lat]} \BibitemShut {NoStop}%
\bibitem [{\citenamefont {Bazavov}\ \emph {et~al.}(2012)\citenamefont {Bazavov}
  \emph {et~al.}}]{Bazavov:2011nk}%
  \BibitemOpen
  \bibfield  {author} {\bibinfo {author} {\bibfnamefont {A.}~\bibnamefont
  {Bazavov}} \emph {et~al.},\ }\bibfield  {title} {\enquote {\bibinfo {title}
  {{The chiral and deconfinement aspects of the QCD transition}},}\ }\href
  {\doibase 10.1103/PhysRevD.85.054503} {\bibfield  {journal} {\bibinfo
  {journal} {Phys. Rev. D}\ }\textbf {\bibinfo {volume} {85}},\ \bibinfo
  {pages} {054503} (\bibinfo {year} {2012})},\ \Eprint
  {http://arxiv.org/abs/1111.1710} {arXiv:1111.1710 [hep-lat]} \BibitemShut
  {NoStop}%
\end{thebibliography}%

\end{document}